\documentclass[a4paper,11pt]{article}
\usepackage{jheppub} 
\usepackage{bbm,bm,graphicx,mathtools,color,hyperref,slashed}
\usepackage{amssymb,amsmath}
\usepackage[T1]{fontenc}

\usepackage{youngtab}

\usepackage{color} 
\usepackage[all]{xy}

\graphicspath{{./Figures/}}

\newcommand{\diff}{\mathrm{d}}
\newcommand{\p}{\partial}

\newcommand{\be}{\begin{equation}}      
\newcommand{\ee}{\end{equation}}     
\newcommand{\beq}{\begin{equation}}      
\newcommand{\eeq}{\end{equation}}     
\newcommand{\bea}{\begin{eqnarray}}      
\newcommand{\eea}{\end{eqnarray}}

\newcommand{\tr}{\mathrm{tr}}

\newcommand{\im}{\mathrm{i}}

\newcommand{\calR}{\mathcal{R}}

\newcommand{\rmc}{\mathrm{c}}
\newcommand{\rme}{\mathrm{e}}
\newcommand{\rmf}{\mathrm{f}}
\newcommand{\rmA}{\mathrm{A}}
\newcommand{\rmF}{\mathrm{F}}
\newcommand{\rmL}{\mathrm{L}}

\newcommand{\rmS}{\mathrm{S}}

\newcommand{\Z}{\mathbbm{Z}}
\newcommand{\Zp}{(\Z_{N+2})_{\psi}}
\newcommand{\Ze}{(\Z_{N+4})_{\eta}}
\def\brc{\langle}
\def\ckt{\rangle}

\def\s{\phantom{=}}

\title{Dynamics from symmetries in chiral $SU(N)$ gauge theories}

\author[a,b]{Stefano Bolognesi,}
\author[a,b]{Kenichi Konishi,}
\author[c,b]{Andrea Luzio}

\affiliation[a]{Department of Physics ``E. Fermi'', University of Pisa, Largo Pontecorvo, 3, Ed. C, 56127, Pisa, Italy, }
\affiliation[b]{INFN, Sezione di Pisa, Largo Pontecorvo, 3, Ed. C, 56127,  Pisa, Italy, }
\affiliation[c]{Scuola Normale Superiore, Piazza dei Cavalieri, 7, 56126,  Pisa, Italy,   }

\emailAdd{stefano.bolognesi@unipi.it}
\emailAdd{kenichi.konishi@unipi.it}
\emailAdd{andrea.luzio@sns.it}

\abstract{
The symmetries and dynamics of  simple  chiral $SU(N)$ gauge theories,  with matter Weyl fermions in a two-index symmetric tensor  and $N+4$ 
anti-fundamental  representations,    are examined, by taking advantage of the recent developments 
involving the ideas of generalized symmetries, gauging of discrete center 1-form symmetries and  mixed 't Hooft anomalies. 
This  class of models are particularly interesting because the conventional 't Hooft anomaly matching constraints allow a chirally symmetric confining vacuum,  with no condensates breaking the $U(1) \times SU(N+4)$ flavor symmetry, and with certain set of massless baryonlike composite fermions saturating all the associated anomaly triangles. Our calculations show that in such a vacuum  the UV-IR matching of some $0$-form$-$$1$-form  
mixed 't Hooft  anomalies fails.  This implies, for the theories with even $N$ at least, that a chirally symmetric 
confining vacuum contemplated earlier in the literature  actually cannot be realized dynamically.  In contrast, a Higgs phase characterized by some 
gauge-noninvariant bifermion condensates passes our improved scrutiny.
}

\begin{document}

\maketitle

\newpage

\section{Introduction}

  In spite of the bulk of knowledge accumulated after almost half-century of studies of vectorlike gauge theories such as $SU(3)$ quantum chromodynamics (QCD), 
  partially based on ever more sophisticated but basically straightforward 
  approximate calculations (lattice simulations),  
as well as  some beautiful theoretical developments in models with  ${\cal N}=1$   or  ${\cal N}=2$ supersymmetries  \cite{AffleckDineSei}-\cite{Cachazo}, \cite{SW,Tachikawa}, 
surprisingly little is known today  about 
 strongly-coupled ordinary (nonsupersymmetric)  chiral gauge theories.  Perhaps it is not senseless to make some more efforts to try to understand this class of theories, which
 Nature might be making use of, in a way as yet unknown to us.    
 
 Such a consideration has led two of us recently to give a systematic look into possible phases of a large class of  chiral gauge theories \cite{BKS,BK}, the first with M. Shifman. 
 To be concrete, we limited ourselves to $SU(N)$ gauge theories with a set of Weyl fermions in a reducible complex representation of $SU(N)$. 
 The gauge interactions in these models become strongly coupled in the infrared.  There are no gauge-invariant bifermion condensates,   no mass  terms or potentials  (of renormalizable type) can be added to deform the theories, including the $\theta$ term,  and the vacuum is unique.   
 
The questions we addressed ourselves to are: 
  (i)  Do these systems confine,  or experience a dynamical Higgs phenomenon (dynamical gauge symmetry breaking)? \,
    (ii)  Do some of them flow into an IR fixed-point CFT? \,
    (iii)  Does the chiral flavor symmetry remain unbroken, or if spontaneously broken, in which pattern?  \,
   (iv)  If there are more than one apparently possible dynamical scenarios,   which one is actually realized in
  the infrared? \,
 (v)  Does the system generate hierarchically disparate mass scales, such as the ones proposed in the "tumbling" scenarios \cite{Raby}?
 and so on.   The general conclusion is that the consideration based on the  't Hooft anomaly matching conditions \cite{th} and on some other consistency conditions
 do restrict the list of possible dynamical scenarios,  but are not sufficiently
  stringent \cite{BKS}-\cite{Goity}.  A more powerful theoretical reasoning is clearly wanted.

Recently the concept of generalized symmetries \cite{AhaSeiTac,GKSW}  has been applied to Yang-Mills theories and QCD like theories, to yield new,  stronger,  version  (involving $0$-form and $1$-form symmetries together) of 't Hooft anomaly matching constraints \cite{GKKS}-\cite{BKL}. 
 The generalized symmetries do not act on local field operators, as in conventional symmetry operations,  but only on extended 
objects, such as closed line or surface operators.\footnote{A familiar example of a  $1$-form symmetry is the   ${\mathbb Z}_N$ center symmetry in $SU(N)$ Yang-Mills theory, acting on 
closed Wilson loops or on Polyakov loops in Eulidean formulation. As is well-known, a vanishing (nonvanishing) VEV of the Polyakov loop can be used as a criterion for detecting confinement (Higgs) phase of the theory.}   The generalized symmetries are all Abelian \cite{AhaSeiTac,GKSW}.  
This last fact was crucial in the recent extension of these new techniques with color $SU(N)$ center ${\mathbb Z}_N$  to theories with fermions in the fundamental representation. The presence of such fermions in the system would normally simply break the center ${\mathbb Z}_N$ symmetry and would prevent us from applying these new techniques.  A color-flavor locking by using appropriate  discrete subgroups of global $U(1)$ symmetries associated with fermion fields, actually allows us to extend the use of $SU(N)$ center ${\mathbb Z}_N$  symmetries in those theories.\footnote{A careful exposition of these ideas can be found e.g., in \cite{Tanizaki}. }

A key  ingredient of these developments is the idea of "gauging a discrete symmetry",  i.e., identifying the field configurations related by the $1$-form (or a higher-form) 
symmetries, and eliminating the consequent redundancies, effectively modifying the path-integral summation rule over gauge fields  \cite{KapSei, Seiberg}.  
Since these generalized symmetries {\it are} symmetries of the models considered, even though they act differently from the conventional ones,  
it is up to us to decide to "gauge"  these symmetries.   Anomalies we encounter in doing so, are indeed obstructions of gauging a symmetry,  i.e.,   a 't Hooft anomaly by definition.   
 And as in the usual application of the 't Hooft anomalies such as the "anomaly matching" between UV and IR theories,  a similar constraint arises 
 in considering the generalized symmetries together with a conventional  ("0-form") symmetry, which has come to be called in recent literature as a "mixed 't Hooft anomaly". 
 Another term of "global inconsistency" was also used to describe a related phenomenon.

In this paper we take a few, simplest chiral gauge theories as exercise grounds, 
and ask whether these new theoretical tools   can be usefully  applied to them, and whether they provide us with new insights into the infrared dynamics and global symmetry realizations of these models.\footnote{ In a recent work we discussed mixed anomalies for a class of chiral gauge theories for which a sub-group of the center of the gauge group does not act on fermions  \cite{BKL}. }

For clarity of presentation, we focus the whole discussion here on a single class of models (${\psi\eta}$ models \cite{BKS,BK}). 
In Sec.~\ref{sec:earlier}  we review the symmetry and earlier results on the possible phases of these theories. In Sec.~\ref{sec:symmetries}  the symmetry group of the systems is discussed more carefully, by taking into account its global aspects. Sec.~\ref{sec:mixedanomalies} and Sec.~\ref{sec:even} contain the derivation of  the anomalies in odd $N$ and even $N$ theories, respectively. In Sec.~\ref{sec:computing} we discuss the UV-IR matching constraints
of certain $0$-form and $1$-form mixed anomalies, and their consequences on the IR dynamics in even $N$ theories. In Sec.~\ref{sec:withoutSZ} the mixed anomalies are reproduced without using the Stora-Zumino descent procedure adopted
 in Sec.~\ref{sec:computing}.
 Summary of our analysis and Discussion are in Sec.~\ref{sec:summary}.   We shall come back to 
more general classes of chiral theories in a separate work.

  \section{The model and  the possible phases  \label{sec:earlier}}

    The  model  we consider in this work  is  an $SU(N)$ gauge theory with Weyl fermions 
\beq
   \psi^{\{ij\}}\,, \quad    \eta_i^B\, , \qquad  ({\footnotesize  i,j = 1,2,\ldots, N\;,\quad B =1,2,\ldots , N+4})\;,
\eeq
in the direct-sum  representation
\be       \yng(2) \oplus   (N+4) \,{\bar   {{\yng(1)}}}\; \ee
of $SU(N)$. This model was studied in  \cite{ACSS,ADS}, \cite{BKS,BK}.\footnote{A recent application of this class of chiral gauge theories is found in \cite{Caccia}.}   This is the simplest of the class of chiral gauge theories known as Bars-Yankielowicz models \cite{BY}. 
The first coefficient of the beta function is
\be   b_0= 11N -  (N+2) - (N+4) = 9N-6\;.  \label{betafn}
\ee
    The fermion kinetic term is given by 
\be
\overline{\psi}\gamma^{\mu}\big(\partial+\calR_{\rmS}(a)\big)_{\mu}P_\rmL\psi+\sum_{B=1}^{N+4}\overline{\eta_{B}}\gamma^{\mu}\big(\partial+\calR_{\rmF^*}(a)\big)_{\mu}P_\rmL\eta_{B}\;,
\ee
with an obvious notation. In order to emphasize that this is the chiral gauge theory, we explicitly write the chiral projector $P_\rmL=\frac{1-\gamma_5}{2}$ in the fermion kinetic terms.   The  symmetry group is
  \be  SU(N)_{\rmc} \times SU(N+4) \times  U(1)_{\psi\eta}\;,   \label{group}
    \ee
    where $U(1)_{\psi\eta}$ is the anomaly-free combination of  $U(1)_{\psi}$ and $U(1)_{\eta}$,
    \be
U(1)_{\psi\eta} : \ \psi\to \rme^{\im (N+4)\alpha}\psi\;, \qquad  \eta \to \rme^{-\im (N+2)\alpha}\eta\;. \qquad \alpha \in \mathbbm{R}
\label{upe0} \;.
\ee

      The group  (\ref{group})  is actually not the true symmetry group of our system, but its covering group.  It captures correctly the local aspects, e.g.,  how the group behaves around the identity element, and thus is sufficient for the consideration of the conventional, perturbative  triangle anomalies associated with it, reviewed below in this section.  
      
      Its global structures however contain some redundancies, which must be modded out appropriately in order to eliminate the double counting. 
They furthermore depend crucially on whether $N$ is odd or even.  These questions will be studied more carefully in Sec.~\ref{sec:symmetries},
as they turn out to be central to the main theme of this work:     the determination of the mixed anomalies and  the associated, generalized 't Hooft anomaly matching conditions.

     \subsection{Chirally symmetric phase
\label{possible0}} 
    
     It was noted earlier  \cite{ACSS,ADS,BKS}  that the standard 't Hooft anomaly matching conditions
     associated with the continuous symmetry group
    $ U(1)_{\psi\eta}  \times  SU(N+4) $
      allowed  a chirally symmetric, confining vacuum in the model.  
    Let us indeed assume that no condensates form, the system confines, and the flavor symmetry is unbroken.
    The candidate massless composite fermions ("baryons")  are:
      \be     {\cal B}^{[AB]}=    \psi^{ij}  \eta_i^A  \eta_j^B \;,\qquad  A,B=1,2, \ldots, N+4\;,\label{baryons10}
\ee
antisymmetric in  $A \leftrightarrow B$.
All the $SU(N+4) \times U(1)_{\psi\eta}$ anomaly triangles  are saturated by  $ {\cal B}^{[AB]}$ as can be seen by inspection of Table~\ref{Simplest0}.
\footnote{There are discrete     unbroken symmetries  ${\mathbbm Z}_{\psi}$ and ${\mathbbm Z}_{\psi}$  which will be defined later (\ref{dp}), (\ref{de}) which are already contained in the covering space (\ref{group}). The discrete anomalies  ${\mathbbm Z}_{\psi}\,  SU(N)^2$, ${\mathbbm Z}_{\psi}\,  SU(N+4)^2$,  ${\mathbbm Z}_{\eta}\,  SU(N)^2$  and    ${\mathbbm Z}_{\eta}\,  SU(N-4)^2$  are also matched as a direct consequence.}
\begin{table}[h!t]
  \centering 
  \begin{tabular}{|c|c|c |c|c|  }
\hline
$ \phantom{{{   {  {\yng(1)}}}}}\!  \! \! \! \! \!\!\!$   & fields  &  $SU(N)_{\rm c}  $    &  $ SU(N+4)$     &   $ {U}(1)_{\psi\eta}   $  \\
 \hline 
  \phantom{\huge i}$ \! \!\!\!\!$  {\rm UV}&  $\psi$   &   $ { \yng(2)} $  &    $  \frac{N(N+1)}{2} \cdot (\cdot) $    & $   N+4$    \\
 & $ \eta^{A}$      &   $  (N+4)  \cdot   {\bar  {\yng(1)}}   $     & $N \, \cdot  \, {\yng(1)}  $     &   $  - (N+2) $ \\
   \hline     
 $ \phantom{ {\bar{   { {\yng(1,1)}}}}}\!  \! \! \! \! \!\!\!$  {\rm IR}&    $ {\cal B}^{[AB]}$      &  $  \frac{(N+4)(N+3)}{2} \cdot ( \cdot )    $         &  $ {\yng(1,1)}$        &    $ -N    $   \\
\hline
\end{tabular}
  \caption{\footnotesize  Chirally symmetric phase of the  $(1,0)$  model.   
The multiplicity, charges and the representation are shown for each  set of fermions. $(\cdot)$ stands for a singlet representation.
}\label{Simplest0}
\end{table}

\subsection{Color-flavor locked Higgs phase   \label{possible1}}

As the theory is very strongly coupled in the infrared (see (\ref{betafn})),  it is also natural to consider the possibility that 
a bifermion condensate 
\be    \brc  \psi^{\{ij\}}   \eta_i^B \ckt =\,   c \,  \Lambda^3   \delta^{j B}\ne 0\;,   \qquad   j, B=1,2,\dots  N\;,   \qquad c \sim O(1) \label{cflocking}
\ee
forms.  $\Lambda$ is the renormailization-invariant scale dynamically generated by the gauge interactions.  The color gauge symmetry is completely (dynamically) broken, leaving however color-flavor diagonal  $SU(N)_{\rm cf} $ symmetry
\beq
SU(N)_{\rm cf} \times  SU(4)_{\rm f}  \times U(1)^{\prime} \,,     \label{hs}
    \eeq
    where $U(1)^{\prime}$ is a combination of $U(1)_{\psi \eta}$ and the elements of $SU(N+4)$ generated by 
\be   \left(\begin{array}{cc}- 2\,   \mathbf{1}_N &  \\ & \frac N2 \mathbf{1}_4\end{array}\right) \;.
\ee
    As (\ref{hs})  is a subgroup of  the original full symmetry group  (\ref{group})
      it can be quite easily verified, by making the decomposition of the fields in the direct sum of representations in  the subgroup,  that a subset of the same baryons  ${\cal B}^{[AB]}$
    saturate all of the triangles associated with the reduced symmetry group.    See Table~\ref{SimplestBis}.  
\begin{table}[h!t]
{
  \centering 
  \begin{tabular}{|c|c|c |c|c|c|  }
\hline
$ \phantom{{{   {  {\yng(1)}}}}}\!  \! \! \! \! \!\!\!$   & fields   &  $SU(N)_{\rm cf}  $    &  $ SU(4)_{\rm f}$     &   $  U(1)^{\prime}   $   &  $({\mathbb{Z}}_{2})_F$    \\
 \hline
   \phantom{\huge i}$ \! \!\!\!\!$  {\rm UV}&  $\psi$   &   $ { \yng(2)} $  &    $  \frac{N(N+1)}{2} \cdot   (\cdot) $    & $   N+4  $   & $1$  \\
 & $ \eta^{A_1}$      &   $  {\bar  {\yng(2)}} \oplus {\bar  {\yng(1,1)}}  $     & $N^2 \, \cdot  \, (\cdot )  $     &   $ - (N+4) $    & $-1$ \\
&  $ \eta^{A_2}$      &   $ 4  \cdot   {\bar  {\yng(1)}}   $     & $N \, \cdot  \, {\yng(1)}  $     &   $ - \frac{N+4}{2}  $   & $-1$  \\
   \hline 
   $ \phantom{{\bar{ \bar  {\bar  {\yng(1,1)}}}}}\!  \! \!\! \! \!  \!\!\!$  {\rm IR}&      $ {\cal B}^{[A_1  B_1]}$      &  $ {\bar  {\yng(1,1)}}   $         &  $  \frac{N(N-1)}{2} \cdot  (\cdot) $        &    $   -(N+4) $     & $-1$ \\
       &   $ {\cal B}^{[A_1 B_2]}$      &  $   4 \cdot {\bar  {\yng(1)}}   $         &  $N \, \cdot  \, {\yng(1)}  $        &    $ - \frac{N+4}{2}$   & $-1$   \\
\hline
\end{tabular}  
  \caption{\footnotesize   Color-flavor locked phase in the $\psi\eta$ model, discussed in Sec.~\ref{possible1}.
  $A_1$ or $B_1$  stand for  $1,2,\ldots, N$,   $A_2$ or $B_2$ the rest of the flavor indices, $N+1, \ldots, N+4$.   The fermion parity   $\psi \to -\psi$, $\eta\to -\eta$ is defined below,  Eq.~(\ref{Z2def}).
   }\label{SimplestBis}
}
\end{table}

\phantom{ciao}

       The low-energy degrees of freedom are  $\tfrac{(N+4)(N+3)}{2}$ massless baryons in  the  first, symmetric phase of  Sec.~\ref{possible0},  
     and $\tfrac{N^2+7N}{2}$ massless baryons together with  $8N+1$  Nambu-Goldstone (NG) bosons, in the second.  They represent physically distinct phases.\footnote{The complementarity  does not work here, as noted in \cite{BKS},  even though the (composite) Higgs scalars    $\psi \eta$  are in the fundamental representation of color.}   The general consensus so far has been that it was not known which of the phases, Sec.~\ref{possible0}, Sec.~\ref{possible1}, or some other phase, was realized in this model.  We shall see below that our analysis based on  the mixed anomalies and generalized 't Hooft anomaly matching constraints
     strongly  favors the dynamical Higgs phase, with bifermion condensate (\ref{cflocking}).  The chirally symmetric phase of Sec.~\ref{possible0}  will be found to be inconsistent.

   \section {Symmetry of the system \label{sec:symmetries}}
   
In this section we examine the symmetry of the system more carefully, taking into account the global aspects of the color and flavor symmetry groups. This is indispensable for the study of the generalized, mixed 't Hooft anomalies, as will be seen below. 

The classical symmetry group of our system is given by 
\bea   G_{\mathrm{class}}
&=& G_{\mathrm{c}} \times G_{\mathrm{f}} \nonumber \\ 
&= &SU(N)_{\mathrm{c}} \times  \frac{U(1)_{\psi} \times U(N+4)_{\eta} }{ \mathbb{Z}_{N}} \;. 
\label{eq:symmetry_classical_psieta}
\eea
The color group is $G_{\mathrm{c}} = SU(N)_{c} $,  and its center acts non-trivially on the matter  fields: 
\be   
 \mathbb{Z}_{N}:  \psi \to  \rme^{  \frac{4\pi \im n }{N}} \psi \;, \qquad     \eta  \to  \rme^{-\frac{ 2 \pi \im n}{N}} \eta  \;, \qquad   n \in \{ 1,\dots, N\} \;. \label{zn}
\ee
The flavor group is $G_{\mathrm{f}} =\frac{U(1)_{\psi} \times U(N+4)_{\eta} }{ \mathbb{Z}_{N}} $. 
The division by $\mathbb{Z}_N$ is understood by the fact that the numerator  overlaps with the center of the gauge group, so this has to be factored out in order to avoid double counting.
Another, equivalent way of writing the flavor part of the classical symmetry group is
\be   G_{\mathrm{f}}=  \frac{U(1)_{\psi} \times U(1)_{\eta} \times  SU(N+4) }{ \mathbb{Z}_{N}  \times \mathbb{Z}_{N+4}} \;. 
\label{scv}
\ee

Quantum mechanically one must consider the effects of the anomalies which reduce the flavor group down to its anomaly-free subgroup. 
This reduction of the symmetry is compactly summarized by the  't~Hooft  instanton effective vertex  
\be
  {\cal L}_{\rm eff}    \sim \rme^{-S_{\mathrm{inst}}} \psi^{N+2} \prod_{B=1}^{N+4} \eta^B\;,\label{thesametHooft}
\ee
 (where the color, spin and spacetime  indices are suppressed) as is well known.  
This vertex explicitly breaks the independent $U(1)$ rotations for $\psi$ and $\eta$. Three different sub-groups left unbroken can be easily seen from (\ref{thesametHooft}). First there is the discrete sub-group of $U(1)_{\psi}$:
\beq
\Zp :  \  \psi \to \rme^{\frac{2\pi \im k }{ N+2}} \psi\;, \; \qquad k \in \{1, \dots, N+2\} \;,   \label{dp} 
\eeq
which leaves $\eta$ invariant. Then there is the discrete sub-group of $U(1)_{\eta}$:
\beq
 \Ze : \    \eta\to \rme^{\frac{2\pi \im p }{ N+4}}\eta\;,   \; \qquad p \in \{1, \dots, N+4\}    \label{de}
\eeq
which leaves $\psi$ invariant. Finally there is a continuous anomaly-free combination of  $U(1)_{\psi}$ and $U(1)_{\eta}$:
\be
U(1)_{\psi\eta} : \ \psi\to \rme^{\im (N+4)\alpha}\psi\;, \qquad  \eta \to \rme^{-\im (N+2)\alpha}\eta\;. \qquad \alpha \in \mathbbm{R}
\label{upe} \;.
\ee
The question that arises now is which is the correct anomaly-free sub-group of $ U(1)_{\psi} \times U(1)_{\eta}$.  Clearly all the three listed above are part of the 
 anomaly-free sub-group, but one must find the minimal description,  in order to avoid the double-counting. It is actually sufficient to consider only $U(1)_{\psi\eta}$ with one of the two discrete group. For example by combining the generator of $\Zp$ with $k=1$ with the element of $U(1)_{\psi\eta}$ with $ \alpha = - \frac{2 \pi}{(N+2)(N+4)}
$  one can obtain the generator of $\Ze$.
But still $U(1)_{\psi\eta} \times \Zp$ contains redundancies.

 From this point on,  we must distinguish the two cases, $N$ odd or   $N$ even.

\subsection {Odd $N$ theories}

\begin{figure}[h!]
\centering
\includegraphics[width=0.35\linewidth]{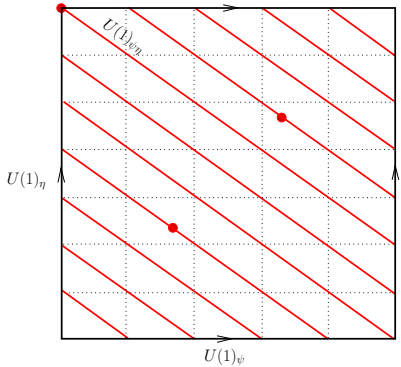}
\caption{\footnotesize The torus $U(1)_{\psi} \times U(1)_{\eta}$ for $N=3$. The edges are identified as the arrows show, the corners represent the identity of the group. The unbroken subgroup $U(1)_{\psi\eta}$ (red line) passing through all the points of the lattice $  (\Z_5)_{\psi} \times (\Z_7)_{\eta}$. The dots indicate  the elements of the center of the gauge group $\Z_3$.}
\label{dispa}
\end{figure}

For odd $N$, the $U(1)_{\psi\eta}$ transformation parameter $\alpha$,  Eq.~(\ref{upe}), exhibits  $2\pi$ periodicity. If we consider the torus $U(1)_{\psi} \times U(1)_{\eta}$, $U(1)_{\psi\eta}$ is a circle that winds $N+4$ times in the $\psi$ direction and $-(N+2)$ times in the $\eta$ direction  before coming back to the origin. See Fig.~\ref{dispa} for the case $N = 3$ where the torus is described as a square with the edges identified, the four corners all correspond to the identity of the group.
Both $\Zp$ and $\Ze$ are sub-groups  of the anomaly-free  $U(1)_{\psi\eta}$. For example by taking $\alpha=\frac{2\pi}{N+2}    \frac{(N+2)+1}{ 2}$ in (\ref{upe}) $\eta$ is left invariant and we recover exactly the generator of $\Zp$. 
The anomaly-free flavor group for odd $N$ is thus:
\be
 G_{\mathrm{f}}=\frac{U(1)_{\psi\eta} \times SU(N+4)} { \mathbb{Z}_{N}\times \mathbb{Z}_{N+4}} \;. 
\label{symmetryNodd}
\ee

The division by $\mathbb{Z}_N$ is due to the fact that the numerator, $U(1)_{\psi\eta}\times SU(N+4)$, overlaps with the center of the gauge group $\mathbb{Z}_N\subset SU(N)$. 
To see this, we ask whether a $U(1)_{\psi\eta}$ transformation   Eq.~(\ref{upe})
can act as the minimal element of ${\mathbb Z}_N\subset SU(N)$:
\be    \psi \to  \rme^{- \frac{4\pi \im}{N}} \psi \;, \qquad     \eta  \to  \rme^{\frac{ 2 \pi \im}{N}} \eta  \;.    \label{arbitrary}
\ee
The solution is
\be   \alpha=  {2\pi\over N}  \frac{N-1}{2}\;,\label{sol}
\ee
as can be easily verified.

The division by $\mathbb{Z}_{N+4}$ can be understood in a similar manner: we consider $U(1)_{\psi\eta}$ with 
\be \alpha=  {2\pi\over N+4}  \frac{N+3}{2}\;,\label{solet}
\ee
 this element acts on fields as 
\be
\psi \to \psi\;, \qquad \eta \to \rme^{-\frac{2\pi \im }{N+4}}\eta\;,     \label{toseethis2}
\ee
which is the center of $SU(N+4)$ flavor symmetry.

The charges of the fields for odd $N$ theory are the same as given 
in Table~\ref{Simplest0}.

\subsubsection{A remark} 

The choice of the generator of  ${\mathbbm Z}_N$, (\ref{arbitrary})  is a little arbitrary. If one required instead
\be    \psi \to  \rme^{\frac{4\pi \im}{N}} \psi \;, \qquad     \eta  \to  \rme^{ -\frac{2 \pi \im}{N}} \eta  \;,    \label{arbitraryBis}
\ee
to be reproduced by $U(1)_{\psi\eta}$
the solution would be 
\be   \alpha=  {2\pi\over N}  \frac{N+1}{2}\;.\label{solBis}   \ee

Similarly for ${\mathbbm Z}_{N+4}$,
\be
\psi \to \psi\;, \qquad \eta \to \rme^{\frac{2\pi i }{N+4}} \,\eta \;,    \label{toseethis3}
\ee
can be reproduced by a $U(1)_{\psi\eta}$ rotation with 
\be \alpha=    {2\pi\over N+4}  \frac{N+5}{2}\;.  \ee

The charges appearing  in (\ref{eq:one}) below would have to be modified accordingly as  
\be   \frac{N-1}{2} \to   \frac{N+1}{2} \;;\qquad \frac{N+3}{ 2} \to  \frac{N+5}{ 2}\;.
\ee
The conclusion of Sec.~\ref{sec:mixedanomalies} below however remains unmodified.

\subsection{Even $N$ theories  \label{note}}

\begin{figure}
\centering
\includegraphics[width=0.35\linewidth]{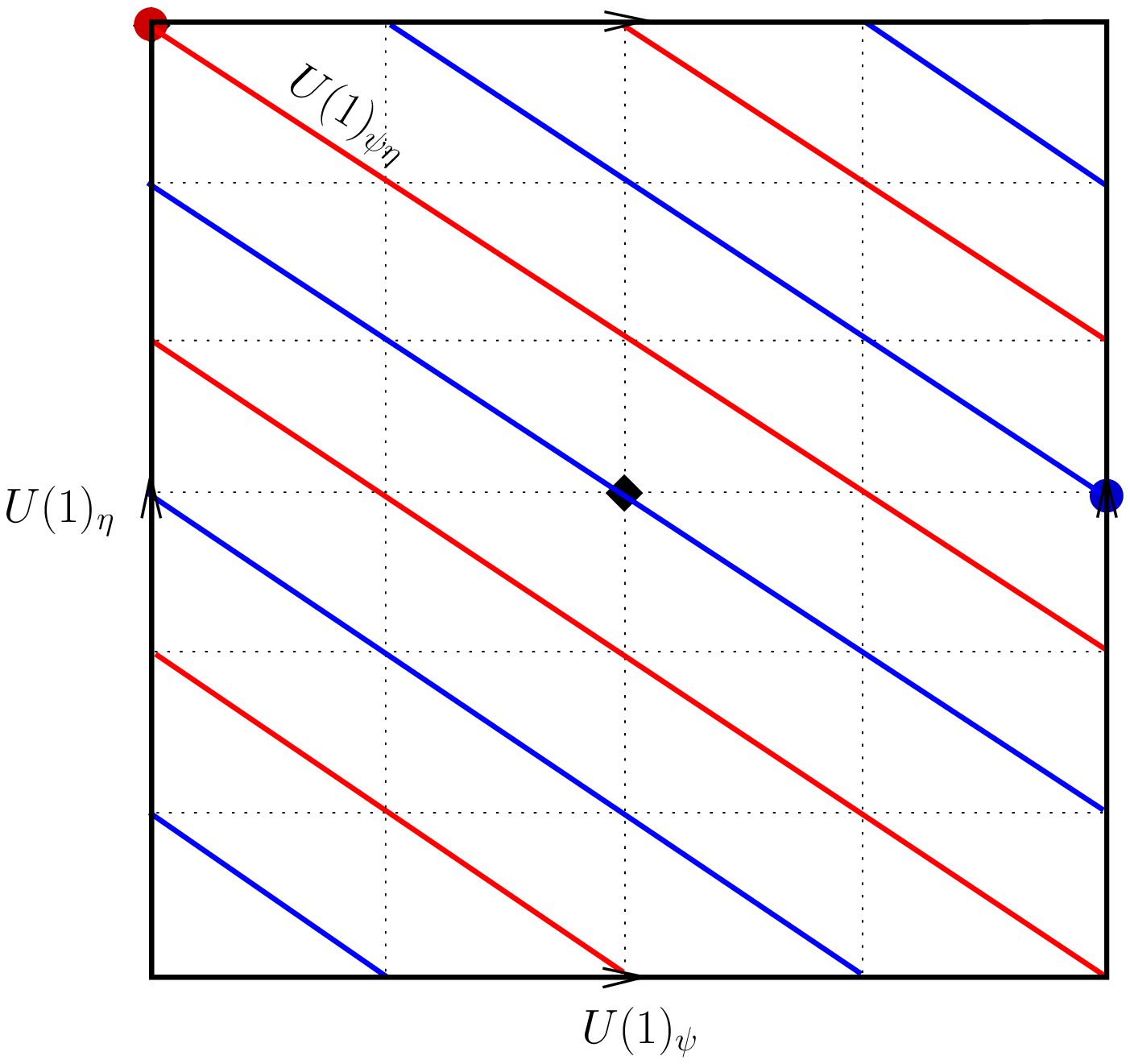} \qquad 
\includegraphics[width=0.35\linewidth]{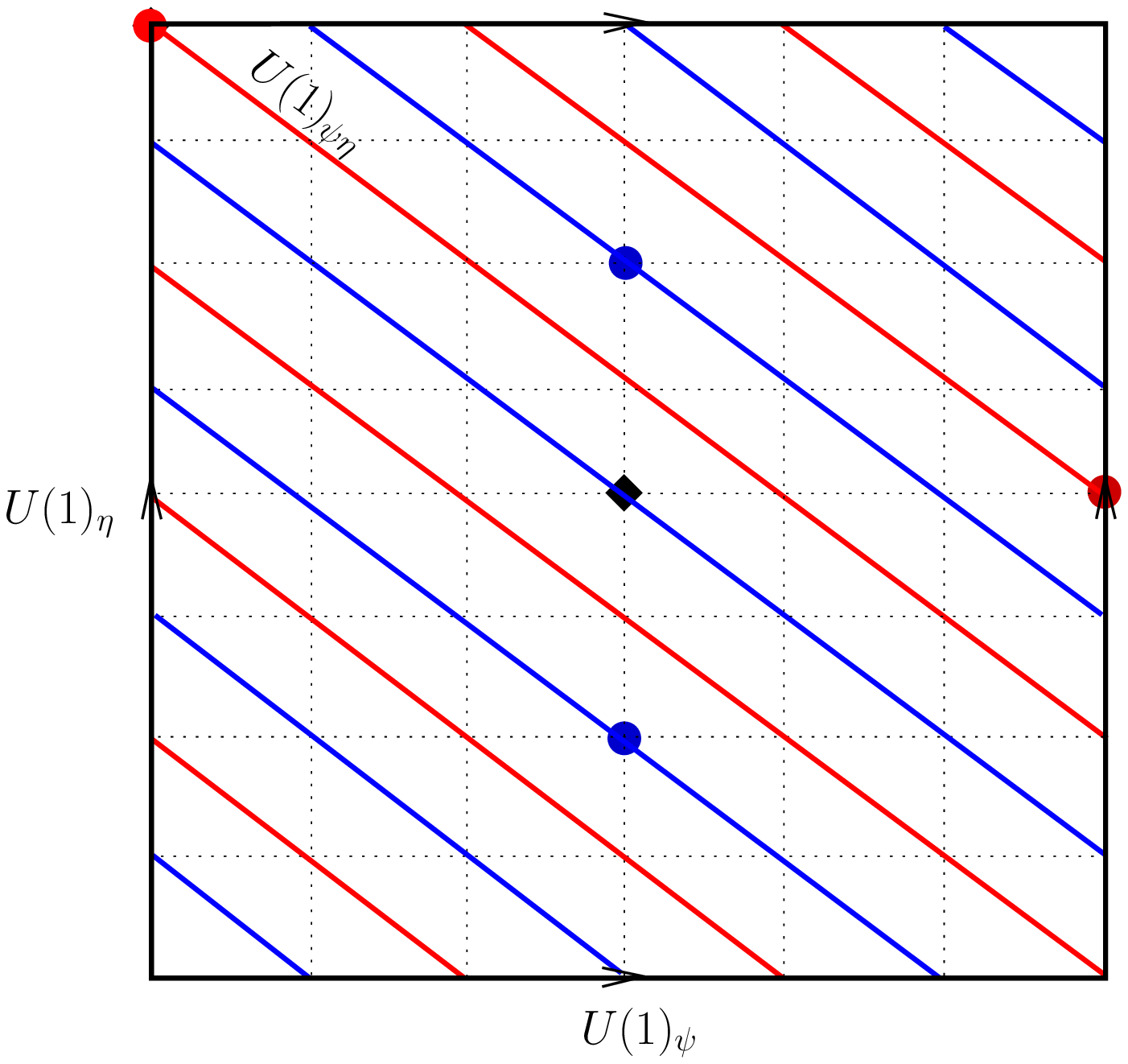}
\caption{\footnotesize  The torus $U(1)_{\psi} \times U(1)_{\eta}$ (for $N=2$ on the left and $N=4$ on the right) and its unbroken subgroup $U(1)_{\psi\eta} \times (\Z_2)_F$ (red line for $U(1)_{\psi\eta} \times \{1\}$ and blue line for $U(1)_{\psi\eta} \times \{-1\}$  ) passing through all the points of the lattice $ \Zp \times \Ze $. The dots indicate the elements of the  group $ (\Z_N)$, diamonds indicate the elements of $(\Z_2)_F$.   $(\mathbb{Z}_2)_F$  is defined below, Eq.~(\ref{Z2def}).}
\label{pari}
\end{figure}

For even $N$, the $U(1)_{\psi\eta}$ transformation parameter $\alpha$,  with the charge convention of (\ref{upe}), exhibits  instead $\pi$ periodicity. It is convenient thus to redefine the $U(1)_{\psi \eta}$ charges as 
\be
\psi\to \rme^{\im { N+4 \over 2}\beta}\psi\;,\;\qquad  \eta\to \rme^{-\im{ N+2 \over 2}\beta }     \label{U1psieta}\eta \;. 
\ee
 With this assignment, the parameter $\beta$ is $2\pi$ periodic. 
$U(1)_{\psi\eta}$ is thus "half" as long as the one for the odd $N$ case;  this is compensated by the fact that now the unbroken sub-group has two disconnected components. See Fig.~\ref{pari} for the cases $N = 2$ and $N=4$.

Let us consider the  fermion parity  defined by 
\be
\psi\to -\psi\;,  \qquad \eta\to -\eta \;,  \label{Z2def}
\ee
which  is equivalent to a $2\pi$ space rotation.
It is clear that $(\mathbb{Z}_2)_F$ is not violated by the 't Hooft  vertex, so let us check if this is not a part of $U(1)_{\psi\eta}$.
 If it were included, there would be  $\beta$ such that 
\be  \rme^{\im { N+4 \over 2}\beta}=\rme^{-\im { N+2 \over 2}\beta}=-1\;. \label{contradicts}\ee
 Multiplying these equations, we get $\rme^{\im \beta}=1$, which  is a contradiction.\footnote{Here we observe a crucial difference with the case of an odd  $N$ theory.   There, the requirement $\rme^{\im {(N+4)}\alpha}=\rme^{-\im {(N+2)}\alpha}=-1$ leads to  $\rme^{2 \im \alpha}=1$,   i.e., $\alpha=0, \pi$, showing that  $(\mathbb{Z}_2)_F  \subset  U(1)_{\psi\eta}$. }

It can be checked that any discrete transformation keeping 't Hooft vertex invariant can be made of $U(1)_{\psi\eta}\times (\mathbb{Z}_2)_F$. For example, $(\mathbb{Z}_{N+2})_\psi$ generated by $\psi\to \rme^{\frac{2\pi\im}{N+2}}\psi$ can also be given by $\big(\beta=\frac{2\pi }{N+2}, -1\big) \in U(1)_{\psi\eta}\times (\mathbb{Z}_2)_F$.   Similarly for $(\mathbb{Z}_{N + 2})_\eta$.

For even $N$,  we thus find that the symmetry group is
\be
G_{\mathrm{f}}=\frac{U(1)_{\psi\eta} \times SU(N+4)  \times (\mathbb{Z}_2)_F} { \mathbb{Z}_{N}\times \mathbb{Z}_{N+4}}\;. 
\label{symmetryNeven}
\ee
The division by $\mathbb{Z}_N$ in Eq.~(\ref{symmetryNeven})  is because the center of the color $SU(N)$ is shared by 
elements in  $U(1)_{\psi\eta} \times  (\mathbb{Z}_2)_F$. Indeed, the gauge transformation with $\rme^{\frac{2\pi\im}{N}}\in \mathbb{Z}_N\subset SU(N)$, 
\be
\psi\to \rme^{\frac{4\pi\im}{N}}\psi\;,\; \qquad \eta \to \rme^{-\frac{2\pi\im}{N}}\eta\;,    \label{ZNpsieta}
\ee
can be written equally well as the following $(\mathbb{Z}_2)_F \times  U(1)_{\psi\eta}  $ transformation:
\be
\psi \to (-1)\, \rme^{\im {N+4\over 2}{2\pi\over N}}\psi = \rme^{-\im {N\over 2}{2\pi\over N}}   \, \rme^{\im {N+4\over 2}{2\pi\over N}}\psi \;, \qquad \eta\to (-1)\, \rme^{-\im {N+2\over 2}{2\pi\over N}}\eta=    \rme^{\im {N\over 2}{2\pi\over N}}     \, \rme^{-\im {N+2\over 2}{2\pi\over N}}\eta\;. \label{ZNequiv}
\ee
Note that the odd elements of $\mathbb{Z}_N$ belong to the disconnected component of $U(1)_{\psi\eta}\times (\mathbb{Z}_2)_F$ while the even elements belong to the identity component.

The division by $\mathbb{Z}_{N+4}$ is understood in a similar manner. The center element $\rme^{\frac{2\pi\im}{N+4}}\in SU(N+4)$ of the flavor group can be identified as the element of $U(1)_{\psi \eta}\times (\mathbb{Z}_2)_F$ as follows:
\be
\psi\to \psi=(-1)\, \rme^{\im {N+4\over 2}{2\pi\over N+4}}\psi=\psi\;,  \qquad \eta\to(-1)\,\rme^{-\im {N+2\over 2}{2\pi\over N+4}}\eta =  \rme^{\im {2\pi\over N+4}}\eta\;. 
\label{ZN+4equiv} \ee
Again, the odd elements of $\mathbb{Z}_{N+4}$ belong to the disconnected component of $U(1)_{\psi\eta} \times (\mathbb{Z}_2)_F$ while the even elements belong to the identity component.

The anomaly-free symmetries and charges for various fields  even $N$ are summarized  in  Table~\ref{symmetriesEvenN}.  
\begin{table}
 \centering 
  \begin{tabular}{|c|c|c|c|c|}
\hline
 fields  &  $SU(N)_{\mathrm{c}}  $    &  $ SU(N+4)$     &   $ U (1)_{\psi \eta}  $  &  $({\mathbb{Z}}_{2})_F$      \\
 \hline
  \phantom{\huge i}$ \! \!\!\!\!$  $\psi$   &   $ { \yng(2)} $  &    $ (\cdot) $    & $   \frac{N+4}{2}$     &  $+1$      \\
 $ \eta$      &   $  \bar{ {\yng(1)}}  $     &  ${ \yng(1)} $   &   $  - \frac{N+2}{2} $  &  $-1$   \\
 \hline  
 $ \phantom{{\bar{  {\bar  {\yng(1,1)}}}}}\!  \! \!\! \! \!  \!\!\!$  $ B^{AB} $      &   $ (\cdot)   $     & $ { \yng(1,1)}$   &    $  - \frac{N}{2} $   &   $-1$     \\
\hline
\end{tabular}
\caption{\footnotesize The charges of various fields with respect to the unbroken symmetry groups for even $N$.  $ B^{AB} $  are  the possible massless composite fermion fields 
discussed in Sec.~\ref{possible0}. The  $({\mathbb{Z}}_{2})_F$  "charge"  in the Table  corresponds  to the transformation 
$\psi \to e^{\im \pi}\psi$,   $\eta \to e^{- \im \pi}\eta$. }
\label{symmetriesEvenN}
\end{table}

\subsection{Symmetry in the Higgs phase}

In the Higgs phase  the group
 (\ref{hs})  is actually a covering space of the true symmetry group which  is given for any  $N$  by
\beq
\frac{SU( N)_{\rm cf} \times  SU(4)_{\rm f}  \times U(1)^{\prime} \times (\mathbb{Z}_2)_F}{\Z_N \times \Z_4} \;,  \label{symmHig}
    \eeq
where  $U(1)'$ has charges given in Table~\ref{SimplestBis}. The fermion parity $ (\mathbb{Z}_2)_F$  
 is left unbroken by the condensate but is not contained in $U(1)'$ so it must be kept in the numerator. The center of $ SU( N)_{\rm cf}$ overlaps completely with  $U(1)'$ so it must be factorized (in fact we may write it as $U(N)_{\rm cf}$). The center of $ SU( 4)_{\rm f}$ also overlaps  with  $U(1)' \times (\mathbb{Z}_2)_F$ which explains the division by $ \Z_4$.

\section{Mixed anomalies:    Odd $N$ case }
\label{sec:mixedanomalies}

 In this section we probe the system with a finer tool, i.e.,  by gauging possible 1-form center symmetries and studying possible mixed 't Hooft anomalies, to see if a stronger constraint emerges. 
 In order to detect the 't~Hooft anomalies, one needs to  introduce the background gauge fields for the global  symmetry $G_{\rmf}$, and check the violation of associated gauge invariance.  Correspondingly to the symmetry of the system, Eq.~(\ref{symmetryNodd}),  we thus introduce
\begin{itemize}
\item $A$: $U(1)_{\psi\eta}$ 1-form gauge field, 
\item $A_\rmf$: $SU(N+4)$ 1-form gauge field, 
\item $B^{(2)}_\rmc$: $\mathbb{Z}_{N}$ 2-form gauge field, 
\item $B^{(2)}_\rmf$: $\mathbb{Z}_{N+4}$ 2-form gauge field. 
\end{itemize}
The field $A = A_{\mu} dx^{\mu}$ gauges the nonanomalous $U(1)_{\psi\eta}$  symmetry discussed in the previous subsection and the field  $A_{\rmf} = A_{\rmf \, \mu} dx^{\mu}$ gauges the  $SU(N+4)$  symmetry.

We recall that in order to gauge a  $\mathbb{Z}_n$ discrete center symmetry of an $SU(n)$ theory,  one introduces a pair of $U(1)$  gauge fields $\big(B^{(2)},B^{(1)}\big)$,  2-form and 1-form fields respectively,   satisfying the constraint \cite{GKSW,GKKS}
\be
n B^{(2)}={\diff} B^{(1)}.    \label{Zn1}
\ee
where $B^{(1)}$  satisfies
\be    \frac{1}{2\pi}   \int_{\Sigma_2}   {\diff} B^{(1)}  =  {\mathbbm Z}\;.
\ee
Existence of the  pair of  gauge fields $\big(B^{(2)},B^{(1)}\big)$  satisfying  relation (\ref{Zn1}) presumes one to have put the system in a topologically nontrivial spacetime $M$.    In such a setting
\be    e^{ \im  \int_{\Sigma_2}   B^{(2)}}    \in  \,   {\mathbbm Z}_n\;  \label{sometimes}
\ee
corresponds to a nontrivial cocycle of $PSU(n) \equiv  \frac{SU(n)}{{\mathbbm Z}_n}$ bundle: an elements of  $w_2(M) \in  H^2(M, {\mathbbm Z}_n)$ known as
 the second  Stiefel-Whitney class.
The constraint   (\ref{Zn1})  satisfies the invariance under the $U(1)$ 1-form gauge transformation, 
\be
B^{(2)}\to B^{(2)}+{\diff} \lambda\;,\quad B^{(1)}\to B^{(1)}+ n \lambda\;.   \label{Zn2}
\ee
The idea is to couple these gauge fields appropriately  to the standard gauge and matter  fields, and  to impose the invariance under the 1-form gauge transformation, Eq.~(\ref{Zn2}), effectively yielding a $PSU(n)$ gauge theory.

This procedure will be applied below both to the color   $SU(N)$  and flavor  $SU(N+4)$ center symmetries.
Actually, the whole analysis of this work could be performed, considering only the gauging of one of the 1-form symmetries,  i.e., ${\mathbbm Z}_N$   or  ${\mathbbm Z}_{N+4}$.  In other words, one may set  $B^{(2)}_\rmf = B^{(1)}_\rmf \equiv  0$,  or  $B^{(2)}_\rmc = B^{(1)}_\rmc \equiv  0$, throughout.  
We are free to choose which one of the 1-form global symmetries, or both, to gauge.   In principle, the implication of our analysis may depend on such a choice.  It turns out, however, that none of the main conclusions of this work (see  Summary in Sec.~\ref{sec:summary}) changes by keeping only one set of   the two-form center gauge fields,     ($B^{(2)}_\rmf, B^{(1)}_\rmf)$, {\it  or}  $(B^{(2)}_\rmc, B^{(1)}_\rmc)$,  but this was not {\it a priori} known.

The $SU(N)$ dynamical gauge field $a$ is embedded into a  $U(N)$ gauge field,
\be
\widetilde{a}=a+{1 \over N}B^{(1)}_\rmc\;, 
\ee
and one requires  invariance under $U(N)$ gauge transformations. 
Similarly, we introduce $U(N+4)$ gauge connection by 
\be
\widetilde{A}_\rmf=A_\rmf+{1\over N+4} B^{(1)}_\rmf\;, 
\ee
and require  $U(N+4)$ gauge invariance instead of the $SU(N+4)$ gauge invariance. 
The pairs of the 1-form$-$2-form $U(1)$ gauge fields are constrained as 
\be   {N}  B^{(2)}_\rmc =  d B^{(1)}_\rmc \;, \qquad    (N+4)    B^{(2)}_\rmf =   d B^{(1)}_\rmf \;.
\ee
The 1-form gauge transformations are defined by 
\bea
&&B^{(2)}_\rmc \to B^{(2)}_\rmc+{\diff} \lambda_\rmc\;,\qquad B^{(1)}_\rmc \to B^{(1)}_\rmc+{N}\lambda_\rmc\;, \nonumber\\
&&B^{(2)}_\rmf \to B^{(2)}_\rmf+{\diff} \lambda_\rmf\;,\qquad  B^{(1)}_\rmf \to B^{(1)}_\rmf+ (N+4)  \lambda_\rmf\;; 
\label{Btrans}
\eea
$\lambda_\rmc$ and $\lambda_\rmf$ are  $U(1)$ gauge fields. 
Under the $\mathbb{Z}_N$ and $\mathbb{Z}_{N+4}$ 1-form transformations,  $U(N)$ and  $U(N+4)$  transform as 
\bea
\widetilde{a}\to \widetilde{a}+\lambda_\rmc\;, \qquad
\widetilde{A}_\rmf\to \widetilde{A}_\rmf+\lambda_\rmf\;. 
\label{eq:one}
\eea
At the same time,  $U(1)_{\psi\eta}$   gauge field is required to  transform as 
\bea
A\to A+{N-1\over 2 }\lambda_\rmc+{N+3\over 2}\lambda_\rmf\;. 
\label{eq:one2}
\eea
The transformation law for  $A$ field is determined by the considerations made around Eqs.~(\ref{sol}) and (\ref{solet}).

In order to have the invariance of the system under the 1-form gauge transformations
the matter fermions must also be appropriately coupled to the 2-form gauge fields. 
Naively, the minimal coupling procedure gives the fermion kinetic term,  
\bea
&&\overline{\psi}\gamma^{\mu}\Big(\partial +\calR_{\rmS}(\widetilde{a})+(N+4)A\Big)_{\mu}P_\rmL\psi  \nonumber\\
&&+ \, \overline{\eta}\gamma^{\mu}\left(\partial + \calR_{\rmF^*}(\widetilde{a})+\widetilde{A}_\rmf-(N+2)A\right)_{\mu}P_\rmL\eta\;. \label{naive0}
\eea
However, this is not invariant under (\ref{eq:one})-(\ref{eq:one2}). Indeed, the above combinations of gauge fields vary as
\bea
\delta\big[\calR_{\rmS}(\widetilde{a})+(N+4)A\big]&=&{N+3\over 2}N\lambda_\rmc+{N+3\over 2}(N+4)\lambda_\rmf\;, \nonumber \\
\delta\big[\calR_{\rmF^*}(\widetilde{a})+\widetilde{A}_\rmf-(N+2)A\big]&=&-{N+1\over 2}N\lambda_\rmc-{N+1\over 2}(N+4)\lambda_\rmf\,. 
\label{noninv} \eea
We therefore require the correct fermion kinetic term with the background gauge fields to be 
\bea
&&\overline{\psi}\gamma^{\mu}\left(\partial+\calR_{\rmS}(\widetilde{a})+(N+4)A-{N+3\over 2}B^{(1)}_\rmc -{N+3\over 2}B^{(1)}_\rmf\right)_{\mu}P_\rmL\psi\nonumber\\
&&+ \, \overline{\eta}\gamma^{\mu}\left(\partial+\calR_{\rmF^*}(\widetilde{a})+\widetilde{A}_\rmf-(N+2)A+{N+1\over 2}B^{(1)}_\rmc+{N+1\over 2}B^{(1)}_\rmf\right)_{\mu}P_\rmL\eta\;.   
\eea

The two-index symmetric fermion $\psi$ feels the gauge field strength
\bea
&&\calR_\rmS\big(  F(\widetilde{a}) \big)+(N+4){\diff} A-{N(N+3)\over 2}B^{(2)}_\rmc-{(N+4)(N+3)\over 2}B^{(2)}_\rmf\nonumber\\
&&=\calR_\rmS\big(F(\tilde{a})-B^{(2)}_\rmc\big)+(N+4)\left[{\diff} A-{N-1\over 2}B^{(2)}_\rmc-{N+3\over 2}B^{(2)}_\rmf\right]\;.     \label{steps0}
\eea
Note that the combination $F(\widetilde{a})-B^{(2)}_\rmc$  is traceless, hence an expression such as  $\calR_\rmS\big(F(\tilde{a})-B^{(2)}_\rmc\big)$   
defined for an $SU(N)$ representation (in this particular case, a symmetric second-rank tensor representation)  is well defined.   Similarly, 
the anti-fundamental fermion $\eta$ feels the gauge field strength 
\bea
&&\calR_{\rmF^*}\big(F(\widetilde{a})\big)+F(\widetilde{A}_\rmf)-(N+2) {\diff} A+{N(N-1)\over 2}B^{(2)}_\rmc+{(N+4)(N+1)\over 2}B^{(2)}_\rmf \qquad \nonumber\\
&&=\calR_{\rmF^*}\big(F(\widetilde{a})-B^{(2)}_\rmc\big)+\big(F(\widetilde{A}_\rmf)-B^{(2)}_\rmf\big)-(N+2)\left[{\diff} A-{N-1\over 2}B^{(2)}_\rmc-{N+3\over 2}B^{(2)}_\rmf\right]\;.  \nonumber\\
\label{etastrength0}
\eea

The low-energy "baryons"   ${\cal B}^{[AB]}$  introduced in Eq.~(\ref{baryons10}) for the chiral symmetric phase are described by the kinetic term,
\be
\overline{\cal B}\,\gamma^{\mu}\left(\p+\calR_{A}(\widetilde{A}_\rmf)  -  {N} A\right)_{\mu}P_\rmL  {\cal B} \;, \label{naiveKTBary11}
\ee
yielding the 1-form gauge invariant form of the field tensor (see Eqs.~(\ref{eq:one})-(\ref{eq:one2})), 
\be   \calR_{A} \big(F(\tilde{A_\rmf})-B^{(2)}_\rmf\big)-   {N}\left[{\diff} A-{N-1\over 2}B^{(2)}_\rmc-{N+3\over 2}B^{(2)}_\rmf\right]  \;.    \label{stepsBaryons}\ee

We are now ready to compute the anomalies following the standard  Stora-Zumino descent  procedure \cite{StoraZumino,Zumino}, as done also in \cite{Tanizaki}.
A good recent review of this renowned procedure can be found in \cite{2groups}.
   The contribution from $\psi$ to the $6D$ Abelian anomaly is 
\bea
&&{1\over 24\pi^2}\, {\tr}_{\calR_\rmS}\left[\left\{\big(F(\tilde{a})-B^{(2)}_\rmc\big)+(N+4)\left[{\diff} A-{N-1\over 2}B^{(2)}_\rmc-{N+3\over 2}B^{(2)}_\rmf\right]
   \right \}^3   \right]    \nonumber\\
&&= \s  {N+4\over 24\pi^2}\, {\tr}\left[\big(F(\tilde{a})-B^{(2)}_\rmc\big)^3\right]\nonumber\\
&& \s +{(N+2)(N+4)\over 8\pi^2}\, {\tr}\left[\big(F(\tilde{a})-B^{(2)}_\rmc\big)^2\right]\wedge \left[{\diff} A-{N-1\over 2}B^{(2)}_\rmc-{N+3\over 2}B^{(2)}_\rmf\right]\nonumber\\
&&\s +{N(N+1)\over 2}{(N+4)^3\over 24\pi^2}\left[{\diff} A-{N-1\over 2}B^{(2)}_\rmc-{N+3\over 2}B^{(2)}_\rmf\right]^3    \;. 
 \label{modified}
\eea
When we write simply "${\tr}$" without an index  the trace is taken in the fundamental representation.
The contribution from   $\eta$ is
\bea
&& {1\over 24\pi^2}\, {\tr}  \left( -[F(\widetilde{a})-B^{(2)}_\rmc]+[F(\widetilde{A}_\rmf)-B^{(2)}_\rmf]-(N+2)\left[{\diff}A-{N-1\over 2}B^{(2)}_\rmc-{N+3\over 2}B^{(2)}_\rmf\right]\right)^3\nonumber\\
&& =    -{(N+4)\over 24\pi^2}\,{\tr}\left[\big(F(\tilde{a})-B^{(2)}_\rmc\big)^3\right]\nonumber\\
&&\s -{(N+2)(N+4)\over 8\pi^2}\,{\tr}\left[\big(F(\tilde{a})-B^{(2)}_\rmc\big)^2\right]\wedge \left[{\diff} A-{N-1\over 2}B^{(2)}_\rmc-{N+3\over 2}B^{(2)}_\rmf\right]\nonumber\\
&&\s +{N\over 24\pi^2}\,{\tr}\left[\big(F(\tilde{A}_\rmf)-B^{(2)}_\rmf\big)^3\right]\nonumber\\
&&\s -{N(N+2)\over 8\pi^2}\,{\tr}\left[\big(F(\tilde{A}_\rmf)-B^{(2)}_\rmf\big)^2\right]\wedge \left[{{\diff}} A-{N-1\over 2}B^{(2)}_\rmc-{N+3\over 2}B^{(2)}_\rmf\right]\nonumber\\
&&\s -{N(N+4)(N+2)^3\over 24\pi^2}\left[{\diff} A-{N-1\over 2}B^{(2)}_\rmc-{N+3\over 2}B^{(2)}_\rmf\right]^3.     \label{2line}
\eea
By summing up these contributions,  we obtain 
\bea
&&{  N\over 24\pi^2}\,{\tr}\left[\big(F(\tilde{A}_\rmf)-B^{(2)}_\rmf\big)^3\right]\nonumber\\
&&-{N(N+2)\over 8\pi^2}\,{\tr}\left[\big(F(\tilde{A}_\rmf)-B^{(2)}_\rmf\big)^2\right]\wedge \left[{\diff} A-{N-1\over 2}B^{(2)}_\rmc-{N+3\over 2}B^{(2)}_\rmf\right]\nonumber\\
&&-{(N+3)(N+4)\over 2}{N^3\over 24\pi^2}\left[{\diff} A-{N-1\over 2}B^{(2)}_\rmc-{N+3\over 2}B^{(2)}_\rmf\right]^3 \;.  \label{result} 
\eea
Note that each factor in the square bracket  in Eqs.~(\ref{modified})-(\ref{result})  is 1-form gauge invariant.

By picking up the boundary  terms  one finds the $5D$ Wess-Zumino-Witten  (WZW) action. 
For instance,   in the limit the 1-form gauging is lifted  (i.e., by setting   $B^{(1)}_\rmf=B^{(1)}_\rmc =0$,  $F(\tilde{A}_\rmf)  \to  F({A}_\rmf)$),  
one recovers, by using the identities 
\be     {\tr} \big(F_{\rmf}^2\big)  =  d \, \Big\{  {\tr}   \big(  A_{\rmf} d A_{\rmf} + \frac{2}{3}  A_{\rmf}^3 \big)\Big\}\;, \quad 
 {\tr} \big(F_{\rmf}^3\big)  =  d\, \Big\{  {\tr}   \big(   A_{\rmf} (d A_{\rmf})^2 + \frac{3}{5}  (A_{\rmf})^5 + \frac{3}{2} A_{\rmf}^3 d A_{\rmf}\big)   \Big\}\;,    
\ee
 the well-known  $5D$ action. The variations of the latter  lead,  by anomaly-inflow,  to the famous   $4D$  
Abelian and nonAbelian anomaly expressions.

Note that 
the dependence on the color gauge field $\widetilde{a}$ disappeared from all terms. 
This is as it should be,  for $N$ odd,  as we are studying the 't Hooft anomaly matching conditions for nonanomalous, {\it continuous}  flavor symmetries.\footnote{Vice versa,  in an even $N$ theory there are anomalies associated with a discrete   ${\mathbbm Z}_2$ symmetry.  The anomaly functionals such as (\ref{psianomaly}) do contain expressions  depending on the color $U(N)_\rmc$ gauge fields   $\widetilde{a}$.  See the discussions below in Sec.~\ref{sec:oldanommatching},  Sec.~\ref{sec:computing}  and Sec.~\ref{sec:summary}. }

As for the  candidate massless "baryons" $\mathcal{B}$   the anomaly functional is given by 
\bea
&&{N+4-4\over 24\pi^2}\,{\tr}\left[\big(F(\tilde{A}_\rmf)-B^{(2)}_\rmf\big)^3\right]\nonumber\\
&&-{N(N+4-2)\over 8\pi^2}\,{\tr}\left[\big(F(\tilde{A}_\rmf)-B^{(2)}_\rmf\big)^2\right]\wedge \left[{\diff} A-{N-1\over 2}B^{(2)}_\rmc-{N+3\over 2}B^{(2)}_\rmf\right]\nonumber\\
&&-{(N+3)(N+4)\over 2}{N^3\over 24\pi^2}\left[{\diff} A-{N-1\over 2}B^{(2)}_\rmc-{N+3\over 2}B^{(2)}_\rmf\right]^3\;,\label{pertmatching}
\eea
as can be seen easily  from Eq.~(\ref{stepsBaryons}).

We are now in the position to compare the anomalies in the UV and IR.       Somewhat surprisingly, we find that 
the IR anomalies Eq.~(\ref{pertmatching})  {\it  exactly}  reproduce   the same  $SU(N+4)\times U(1)_{\psi\eta}$  't~Hooft anomalies   of the UV theory Eq.~(\ref{result}),   independently of whether  or not 
 the 2-form gauge fields  $\big( B^{(2)}_\rmc, B^{(2)}_\rmf \big)$ are introduced!   
 
  Actually,  this is a simple consequence of the fact  that  without the 1-form gauging, these anomalies matched in the UV and IR (the earlier observation, see Sec.~\ref{possible0}).  The coefficients in various triangle diagrams involving  $SU(N+4)$ and $U(1)_{\psi\eta}$ vertices, 
  computed by using the UV and IR fermion degrees of freedom,  are equal.  Upon gauging the 
  1-form center symmetries, the  external  $SU(N+4)$ and $U(1)_{\psi \eta}$ gauge 
fields  are replaced  by the center-1-form-gauge-invariant combinations,  both in the UV and IR, 
as in Eq.~(\ref{steps0}), Eq.~(\ref{etastrength0}), Eq.~(\ref{stepsBaryons}),   but clearly  the  UV-IR matching of various anomalies continue to hold.  It turns out that the situation is different when the UV-IR anomaly matching involves a discrete symmetry, as in even $N$ theories discussed below.  See below.

\section{Mixed anomalies: Even $N$ case \label{sec:even}}

We discuss now  the even $N$ theories.   The calculation of the anomalies, 1-form gauging and anomaly matching checks go through mostly  as in the odd $N$ case discussed above, by taking into account appropriately the difference in the $U(1)_{\psi\eta}$  charges of the matter fields and in the center symmetries themselves, as well as the presence of an independent discrete $({\mathbb Z}_{2})_{F}$  symmetry.  However  the  
conclusion turns out to be  qualitatively different.

\subsection{Calculation of anomalies}

To detect the anomalies of global symmetry $G_{\rmf}$, Eq.~(\ref{symmetryNeven}),    we introduce the gauge fields 
\begin{itemize}
\item $A$: $U(1)_{\psi\eta}$ 1-form gauge field, 
\item $A_2^{(1)}$:  $({\mathbb Z}_{2})_{F}$  1-form gauge field,
\item $A_\rmf$: $SU(N+4)$ 1-form gauge field, 
\item $B^{(2)}_\rmc$: $\mathbb{Z}_{N}$ 2-form gauge field, 
\item $B^{(2)}_\rmf$: $\mathbb{Z}_{N+4}$ 2-form gauge field.
\end{itemize}
$({\mathbb Z}_{2})_{F}$  is an ordinary ($0$-form) discrete symmetry, and we introduced accordingly a 1-form gauge field 
\be  A_2^{(1)}\;,  \qquad   \delta  A_2^{(1)}  =  \frac{1}{2}  \,   \diff \,   \delta A_2^{(0)}\;. \ee
The  $({\mathbb Z}_{2})_{F}$  variation in the $4D$ action is described by, 
\be      \delta  A_2^{(0)} = \pm 2 \pi\;, \qquad {\rm i.e.} \;, \qquad   \psi \to   e^{\im \pi} \psi=-\psi, \quad \eta \to  e^{-\im \pi} \eta= -\eta\;.  \label{Fparity}
\ee
In order to avoid misunderstandings, let us repeat that  $A_2^{(1)}$  is a gauge field formally introduced to describe an ordinary ($0$-form) $({\mathbb Z}_{2})_{F}$   symmetry. In this sense it is perfectly analogous to the 
$U(1)_{\psi\eta}$  gauge field, $A$.   $\big(B_\rmc^{(2)}$, $B_\rmf^{(2)}\big)$ are instead introduced to "gauge" the 1-form center ($\mathbb{Z}_N$  and $\mathbb{Z}_{N+4}$) symmetries  \footnote{In order to completely dispel the risk of confusion, it might have been a good idea to put suffix such as in
$({\mathbb Z}_{2})_{F}^{(0)}$,  $\mathbb{Z}_N^{(1)}$, or  $\mathbb{Z}_{N+4}^{(1)}$, to show explicitly which types of symmetry we are talking about.    We refrained ourselves from doing so in this work, however, in order to avoid cluttered formulae,   
and confiding in the  attentiveness of the reader.  Another reason is that the symbol $\mathbb{Z}_N$, e.g.,  is used both to indicate the particular symmetry type and to stand for the cyclic group $\mathbb{C}_N$ itself. 
}. The procedure was reviewed briefly at the beginning of Sec.~\ref{sec:mixedanomalies},  in the case of  odd $N$ theories. 

For even $N$ theories under consideration here,   the construction is similar.   We introduce two pairs of  gauge fields  $\big(B_\rmc^{(2)}$, $B_\rmc^{(1)}\big)$  and $\big(B_\rmf^{(2)}$, $B_\rmf^{(1)}\big)$, satisfying the constraints \footnote{See the discussion at the beginning of Sec.~\ref{sec:mixedanomalies} for the meaning of these constraints. }
\be  N  B_\rmc^{(2)} = d  B_\rmc^{(1)}\;; \qquad  (N+4)  B_\rmf^{(2)} = d   B_\rmf^{(1)}\;.    \label{1fgconstraint}
\ee
Under the gauged (1-form) center transformations, these fields transform as
\be
B_\rmc^{(2)}\to B_\rmc^{(2)}+\diff \lambda_\rmc\;,  \qquad  B_\rmc^{(1)}\to B_\rmc^{(1)}+ N  \lambda_\rmc \;,      \label{ZN1FG}
\ee
\be
B_\rmf^{(2)}\to B_\rmf^{(2)}+\diff \lambda_\rmf\;,  \qquad    B_\rmf^{(1)}\to B_\rmf^{(1)}+(N+4) \lambda_\rmf\;,  \label{ZN41FG}
\ee
which respect  the constraints (\ref{1fgconstraint}).  Now the whole system must be made invariant under these transformations, and this requires the
 gauge fields  $A$, $A_2$, $A_\rmf$,  color $SU(N)$ gauge field $a$,  as well as the fermions,  be all coupled  appropriately to   $\big(B_\rmc^{(2)}$, $B_\rmc^{(1)}\big)$  and $\big(B_\rmf^{(2)}$, $B_\rmf^{(1)}\big)$ fields.

To achieve this 
we first embed the dynamical $SU(N)$ gauge field $a$ into a  $U(N)$ gauge field $\widetilde{a}$ as 
\be
\widetilde{a}=a+{1\over N}B^{(1)}_\rmc,
\ee
and the $SU(N+4)$ flavor gauge field as $U(N+4)$ gauge field $\widetilde{A}_\rmf$ as 
\be
\widetilde{A}_\rmf=A_\rmf+{1\over N+4}B^{(1)}_\rmf\,. 
\ee
Under the center of $SU(N)$, the symmetry-group element $(\rme^{\im \alpha}, (-1)^n, g_\rmf)\in U(1)\times \mathbb{Z}_2\times SU(N+4)$ is identified as 
(see Eq.~(\ref{ZNequiv}))
\be
(\rme^{\im \alpha}, (-1)^n, \, g_\rmf)\sim (\rme^{\im (\alpha-{2\pi\over N})}, (-1)^n\rme^{\im{2\pi\over N}{N\over 2}},  \, g_\rmf)\,. \label{equiv11}
\ee
This means that $U(1)_{\psi \eta}$ gauge field $A$ has charge $-1$, $(\mathbb{Z}_2)_F$ gauge field $A_2^{(1)}$ has charge ${N\over 2}$, and $U(N+4)$ gauge field $\widetilde{A}_\rmf$ has charge $0$ under the $U(1)$ 1-form gauge transformation $\lambda_\rmc$ for $B_\rmc^{(2)}$. 

Similarly, the division by $\mathbb{Z}_{N+4}$ means that we identify  (see Eq.~(\ref{ZN+4equiv}))
\be
(\rme^{\im \alpha}, (-1)^n, \,g_\rmf)\sim (\rme^{\im (\alpha-{2\pi\over N+4})},  (-1)^n\rme^{\im {N+4\over 2}{2\pi\over N+4}},  \,g_\rmf\,\rme^{2\pi\im\over N+4})\;, 
\label{redundant}  \ee
and this determines the charges under $\lambda_\rmf$.

These considerations determine uniquely the way  the 1-form gauge fields transform  under (\ref{ZN1FG}) and (\ref{ZN41FG}): 
\bea
&&\widetilde{a}\to \widetilde{a}+\lambda_\rmc\;,  \nonumber  \\
&&A\to A-\lambda_\rmc-\lambda_\rmf\;,   \nonumber    \\
&&A_2^{(1)}\to A_{2}^{(1)}+{N\over 2}\lambda_\rmc+{N+4\over 2}\lambda_\rmf  \label{eq:1form_trans_Z2}\;, \nonumber   \\
&&\widetilde{A}_\rmf\to \widetilde{A}_\rmf+\lambda_\rmf\;.   \label{werequire} 
\eea

The crucial ingredient in our analysis now is the  nontrivial 't Hooft fluxes carried by the   
 (${\mathbbm Z}_N$ and ${\mathbbm Z}_{N+4}$) 2-form gauge fields  $B_\rmc^{(2)}$ and  $B_\rmf^{(2)}$,    
 \be      \frac{1}{2\pi} \int_{\Sigma_2}    B_\rmc^{(2)}    =     \frac{ n_1 }{N}\;,    \qquad     n_1 \in   {\mathbbm Z}_N\;,   \label{Byconstruction2}
 \ee
 \be      \frac{1}{2\pi} \int_{\Sigma_2}   B_\rmf^{(2)}    =     \frac{
 m_1 }{N+4}\;,   \qquad     m_1  \in   {\mathbbm Z}_{N+4}\;,   \label{Byconstruction222}
 \ee
 in a closed two-dimensionl space,    ${\Sigma_2}$.    On topologically nontrivial four dimensional spacetime
 of Euclidean signature  ${\Sigma_2} \times {\Sigma_2}$    one has then 
  \be      \frac{1}{8\pi^2} \int_{\Sigma_4}   (B_\rmc^{(2)})^2   =   \frac{n }{N^2}\;, \qquad     \frac{1}{8 \pi^2} \int_{\Sigma_4}   (B_\rmf^{(2)})^2 =  \frac{
 m }{(N+4)^2}\;, \label{Byconstruction4}
 \ee
   where $n \in   {\mathbbm Z}_N$ and     $m  \in   {\mathbbm Z}_{N+4}$,  and   an extra factor $2$ with respect to (\ref{Byconstruction2})  is due to the two possible ways   the two $B_\rmc^{(2)}$ fields 
 are distributed on the two ${\Sigma_2}$'s  (similarly for   $B_\rmf^{(2)}$).

The fermion kinetic term with the background gauge field is obtained by the minimal coupling procedure as 
\bea
&&\overline{\psi}\gamma^{\mu}\left(\partial +\calR_{\rmS}(\widetilde{a})+{N+4\over 2}A+A_2\right)_{\mu}P_\rmL\psi\;  \nonumber\\
&&+\,\overline{\eta}\gamma^{\mu}\left(\partial +\calR_{\rmF^*}(\widetilde{a})+\widetilde{A}_\rmf-{N+2\over 2}A-A_2\right)_{\mu}P_\rmL\eta\;. \label{naive}
\eea
Here, $A_2$ represents the coupling to the fermion parity $(-1)^F$, so its coefficient is meaningful only modulo $2$, and we fix the convention here.\footnote{If the ${\mathbbm Z}_2$ charges  were assigned as $(+1, +1)$, rather than  $(+1, -1)$, as in Eq.~(\ref{naive}),  some coefficients in Eq.~(\ref{UVanomaly})
would change, but the final results would not change. } 
 With this assignment of charges, each covariant derivative turns out to be invariant under 1-form gauge transformations without introducing extra terms.  This is of course a direct reflection of the equivalence, (\ref{ZNpsieta}) and (\ref{ZNequiv}), 
 or  (\ref{equiv11}), (\ref{redundant}), i.e., of  the requirement that the ${\mathbbm Z}_N\subset SU(N)$ transformation is canceled by 
 $U(1)_{\psi\eta} \times {\mathbbm Z}_2$  (and similarly for the ${\mathbbm Z}_{N+4}$ symmetry).

We  compute the anomalies again  by applying the Stora-Zumino descent procedure starting with a $6D$ anomaly functional. 
The two-index symmetric fermion $\psi$ feels the gauge field strength
\bea
\calR_\rmS\big(  F(\widetilde{a}) \big)+{N+4\over 2}{\diff} A+\diff A_2
&=&     \calR_\rmS\big(F(\tilde{a})-B^{(2)}_\rmc\big)+{N+4\over 2}\left[{\diff} A+B^{(2)}_\rmc+B^{(2)}_\rmf\right]\nonumber\\
&&+\left[\diff A_2^{(1)}-{N\over 2}B^{(2)}_\rmc-{N+4\over 2}B^{(2)}_\rmf\right]\;,     \label{steps}
\eea
where appropriate 2-form gauge fields have been introduced so that each term is now 1-form gauge invariant.  
Similarly,  the anti-fundamental fermion $\eta$ feels the gauge field strength 
\bea
&&\calR_{\rmF^*}\big(  F(\widetilde{a}) \big)+F(\widetilde{A}_\rmf)-{N+2\over 2} {\diff} A-\diff A_2  \nonumber\\
&& = -[F(\widetilde{a})-B^{(2)}_\rmc]+[F(\widetilde{A}_\rmf)-B^{(2)}_\rmf]-{N+2\over 2} \left[{\diff} A+B^{(2)}_\rmc+B^{(2)}_\rmf\right]\nonumber\\
&&\s -\left[\diff A_2^{(1)}-{N\over 2}B^{(2)}_\rmc-{N+4\over 2}B^{(2)}_\rmf\right]\;. 
\label{etastrength}
\eea

The low energy "baryons"  gives
\bea
&& \calR_{\rmA}\big(F(\widetilde{A}_\rmf)\big)-{N\over 2} {\diff} A-\diff A_2  \nonumber  \\   
&&=  \calR_{\rmA}  [F(\widetilde{A}_\rmf)-B^{(2)}_\rmf]-{N\over 2} \left[{\diff} A+B^{(2)}_\rmc+B^{(2)}_\rmf\right] 
-\left[\diff A_2^{(1)}-{N\over 2}B^{(2)}_\rmc-{N+4\over 2}B^{(2)}_\rmf\right]\;.  \nonumber  \\  
\label{baryonstrength}
\eea

Before proceeding to the calculation, let us make a brief  pause.  We have already noted that in contrast to the odd $N$ systems considered in Sec.~\ref{sec:mixedanomalies}, 
the fermion kinetic terms  in an even $N$ theory (\ref{naive})  are  invariant under  the center gauge transformations, Eq.~(\ref{ZN1FG}),  Eq.~(\ref{ZN41FG}),  Eq.~(\ref{werequire}), without explicit addition of terms involving  $B^{(2)}_\rmc$ and $B^{(2)}_\rmf$  (cfr. see  Eq.~(\ref{noninv}) for the odd $N$ case).  Thus the rewriting made above (\ref{steps})-(\ref{baryonstrength}) might look redundant at first sight:  these expressions
appear to be actually independent of $B^{(2)}_\rmc$ and $B^{(2)}_\rmf$.  This, however, is not quite correct.   
If one were to proceed with calculation without making each term 1-form gauge invariant, as done above,  the resulting anomaly expressions would not be invariant under the 1-form  ($\mathbb{Z}_N$  and $\mathbb{Z}_{N+4}$)  center gauge transformations, so that there would be no guarantee that the mixed anomalies have been correctly evaluated in the reduced  $PSU(N) $  or $PSU(N+4) $ theories.
We thus prefer to work with explicitly 1-form gauge invariant forms at each step of the calculation below.\footnote{In the standard anomaly calculation in $4D$ \`a la Fujikawa (Sec.~\ref{sec:withoutSZ}),  the introduction of these center gauge fields are seen more straightforwardly as a modification of the theory.  
}


Let us proceed to the $6D$ anomaly functionals due to these fermions:  $\psi$ gives, from Eq.~(\ref{steps}),\footnote{Actually,  $B^{(2)}_\rmf$  (but not  
$B^{(2)}_\rmc$!)   drops out completely from the expression below (\ref{psianomaly}),  as can be seen from the first line. This is correct, as  $\psi$ is a singlet of $SU(N+4)$ and consequently  Eq.~(\ref{steps}) does not contain the 
$SU(N+4)$ gauge fields. This can be used as a check of the calculations below. }
\bea  && {1\over 24\pi^2}\, {\tr}    \left(   \calR_\rmS\big(F(\tilde{a})-B^{(2)}_\rmc\big)+{N+4\over 2}\left[{\diff} A+B^{(2)}_\rmc+B^{(2)}_\rmf\right]   \right.  \nonumber \\ && \qquad \qquad  \ \left.      +\left[\diff A_2^{(1)}-{N\over 2}B^{(2)}_\rmc-{N+4\over 2}B^{(2)}_\rmf\right]  \right)^3   \nonumber \\
&& = \s {(N+4)\over 24\pi^2}\, \tr\left[\big(F(\tilde{a})-B^{(2)}_\rmc\big)^3\right]   \nonumber \\
&& \s + {(N+2)(N+4)\over 16  \pi^2}\, \tr\left[\big(F(\tilde{a})-B^{(2)}_\rmc\big)^2\right]\wedge \left[{\diff} A+B^{(2)}_\rmc+B^{(2)}_\rmf\right] \nonumber\\
&& \s +  \frac{N(N+1)}{2\cdot 24 \pi^2} \left(\frac{N+4}{2}\right)^3  \left[{\diff} A+B^{(2)}_\rmc+B^{(2)}_\rmf\right]^3 \nonumber \\
&&  \s  +  \frac{N+2}{8 \pi^2}  \, {\tr}  \big(F(\tilde{a})-B^{(2)}_\rmc\big)^2   \left[\diff A_2^{(1)}-{N\over 2}B^{(2)}_\rmc-{N+4\over 2}B^{(2)}_\rmf\right]    \nonumber \\
&& \s  +  \frac{1}{8\pi^2}  \left(\frac{N+4}{2}\right)^2 \frac{N(N+1)}{2}   \left[{\diff} A+B^{(2)}_\rmc+B^{(2)}_\rmf\right]^2    \left[\diff A_2^{(1)}-{N\over 2}B^{(2)}_\rmc-{N+4\over 2}B^{(2)}_\rmf\right]   \nonumber \\
&&  \s   +      \frac{1}{8\pi^2}  \left(\frac{N+4}{2}\right) \frac{N(N+1)}{2}   \left[{\diff} A+B^{(2)}_\rmc+B^{(2)}_\rmf\right]    \left[\diff A_2^{(1)}-{N\over 2}B^{(2)}_\rmc-{N+4\over 2}B^{(2)}_\rmf\right]^2            \nonumber \\
&& \s   +   {1 \over 24\pi^2} \frac{N(N+1)}{2}\left[\diff A_2^{(1)}-{N\over 2}B^{(2)}_\rmc-{N+4\over 2}B^{(2)}_\rmf\right]^3 \;. \label{psianomaly}
\eea
The contribution of $\eta$ is   (from Eq.~(\ref{etastrength})):
\bea     && {1\over 24\pi^2}\, {\tr} \, {\huge\{ }-[F(\widetilde{a})-B^{(2)}_\rmc]+[F(\widetilde{A}_\rmf)-B^{(2)}_\rmf]-{N+2\over 2} \left[{\diff} A+B^{(2)}_\rmc+B^{(2)}_\rmf\right] 
\nonumber \\
&&      -\left[\diff A_2^{(1)}-{N\over 2}B^{(2)}_\rmc-{N+4\over 2}B^{(2)}_\rmf\right] 
 {\huge\}}^3   \nonumber \\
 &&  =  -  {(N+4)\over 24\pi^2} \,\tr\left[\big(F(\tilde{a})-B^{(2)}_\rmc\big)^3\right]    +      {N\over 24\pi^2}\tr \left[\big(F(\tilde{A}_\rmf)-B^{(2)}_\rmf\big)^3\right]   \nonumber \\
&&\s   - {(N+2)(N+4)\over 16  \pi^2}\,\tr\left[\big(F(\tilde{a})-B^{(2)}_\rmc\big)^2\right]\wedge \left[{\diff} A+B^{(2)}_\rmc+B^{(2)}_\rmf\right] \nonumber\\
&&\s    -     \frac{N}{8 \pi^2} \frac{N+2}{2} \,{\tr}  [F(\widetilde{A}_\rmf)-B^{(2)}_\rmf]^2  \left[{\diff} A+B^{(2)}_\rmc+B^{(2)}_\rmf\right]    \nonumber \\
&& \s   -{N(N+4)(N+2)^3\over  8 \cdot 24\pi^2}  \left[{\diff} A+B^{(2)}_\rmc+B^{(2)}_\rmf\right]^3 \nonumber \\
&& \s   -     \frac{N+4}{8 \pi^2}\,  {\tr}  \big(F(\tilde{a})-B^{(2)}_\rmc\big)^2   \left[\diff A_2^{(1)}-{N\over 2}B^{(2)}_\rmc-{N+4\over 2}B^{(2)}_\rmf\right]    \nonumber \\
&& \s   -     \frac{N}{8 \pi^2}  \,{\tr}  [F(\widetilde{A}_\rmf)-B^{(2)}_\rmf]^2  \left[\diff A_2^{(1)}-{N\over 2}B^{(2)}_\rmc-{N+4\over 2}B^{(2)}_\rmf\right]    \nonumber \\
&&\s   -  \frac{1}{8\pi^2}  \left(\frac{N+2}{2}\right)^2    N (N+4)  \left[{\diff} A+B^{(2)}_\rmc+B^{(2)}_\rmf\right]^2    \left[\diff A_2^{(1)}-{N\over 2}B^{(2)}_\rmc-{N+4\over 2}B^{(2)}_\rmf\right]   \nonumber \\
&& \s   -         \frac{1}{8\pi^2}  \left(\frac{N+2}{2}\right) N  (N+4)  \left[{\diff} A+B^{(2)}_\rmc+B^{(2)}_\rmf\right]    \left[\diff A_2^{(1)}-{N\over 2}B^{(2)}_\rmc-{N+4\over 2}B^{(2)}_\rmf\right]^2            \nonumber \\
&&\s   -  {1 \over 24\pi^2} N (N+4) \left[\diff A_2^{(1)}-{N\over 2}B^{(2)}_\rmc-{N+4\over 2}B^{(2)}_\rmf\right]^3 \;.  \label{etaanomaly}
\eea  
The sum of the UV anomalies is
\bea  &&   +      {N\over 24\pi^2}\tr \left[\big(F(\tilde{A}_\rmf)-B^{(2)}_\rmf\big)^3\right]\nonumber\\
&&  -     \frac{N}{8 \pi^2} \frac{N+2}{2} {\tr}  [F(\widetilde{A}_\rmf)-B^{(2)}_\rmf]^2   \left[{\diff} A+B^{(2)}_\rmc+B^{(2)}_\rmf\right]    \nonumber \\
&&  -{N^3 (N+4)(N+3) \over  16 \cdot 24\pi^2}  \left[{\diff} A+B^{(2)}_\rmc+B^{(2)}_\rmf\right]^3 \nonumber \\
&&  -     \frac{2}{8 \pi^2}  {\tr}  \big(F(\tilde{a})-B^{(2)}_\rmc\big)^2   \left[\diff A_2^{(1)}-{N\over 2}B^{(2)}_\rmc-{N+4\over 2}B^{(2)}_\rmf\right]     \nonumber \\
&&  -     \frac{N}{8 \pi^2}  {\tr}  [F(\widetilde{A}_\rmf)-B^{(2)}_\rmf]^2  \left[\diff A_2^{(1)}-{N\over 2}B^{(2)}_\rmc-{N+4\over 2}B^{(2)}_\rmf\right]    \nonumber \\
&& -  \frac{1}{8\pi^2} \frac{N(N+4)(N^2+3N+4)}{8}      \left[{\diff} A+B^{(2)}_\rmc+B^{(2)}_\rmf\right]^2    \left[\diff A_2^{(1)}-{N\over 2}B^{(2)}_\rmc-{N+4\over 2}B^{(2)}_\rmf\right]   \nonumber \\
&&   -       \frac{1}{8\pi^2}  \frac{N(N+3)(N+4)}{4}   \left[{\diff} A+B^{(2)}_\rmc+B^{(2)}_\rmf\right]    \left[\diff A_2^{(1)}-{N\over 2}B^{(2)}_\rmc-{N+4\over 2}B^{(2)}_\rmf\right]^2     \nonumber \\
&&  -  {1 \over 24\pi^2} \frac{N(N+7)}{2}  \left[\diff A_2^{(1)}-{N\over 2}B^{(2)}_\rmc-{N+4\over 2}B^{(2)}_\rmf \right]^3      \;.
   \label{UVanomaly}
\eea  
In the IR, the "baryons"  Eq.~(\ref{baryons10}) yield, from Eq.~(\ref{baryonstrength}),    the $6D$ anomaly\footnote{Note  that  $B^{(2)}_\rmc$  actually  drops out completely from this expression,  as is clear from the first line.
            This is as it should be, as the baryons are color  $SU(N)$  singlets:  they are coupled neither to $SU(N)$ gauge fields nor to  
            ${\mathbbm Z}_N$   gauge fields $B^{(2)}_\rmc$. 
               This can again be used as a check in the following calculations.}
\bea && {1\over 24\pi^2}\, {\tr}    \left(   \calR_\rmA (F(\widetilde{A}_\rmf)-B^{(2)}_\rmf )- {N\over 2}\left[{\diff} A+B^{(2)}_\rmc+B^{(2)}_\rmf\right]   \right.  \nonumber \\ && \qquad \qquad  \ \left.    -  \left[\diff A_2^{(1)}-{N\over 2}B^{(2)}_\rmc-{N+4\over 2}B^{(2)}_\rmf\right]  \right)^3       \nonumber \\
&&  =   \s      {N+4-4  \over 24\pi^2}\tr \left[\big(F(\tilde{A}_\rmf)-B^{(2)}_\rmf\big)^3\right]\nonumber\\
&&  \s  -{N^3 (N+4)(N+3) \over  16 \cdot 24\pi^2}  \left[{\diff} A+B^{(2)}_\rmc+B^{(2)}_\rmf\right]^3 \nonumber \\
&& \s  -     \frac{N+2}{8 \pi^2} \frac{N}{2} {\tr}  [F(\widetilde{A}_\rmf)-B^{(2)}_\rmf]^2   \left[{\diff} A+B^{(2)}_\rmc+B^{(2)}_\rmf\right]    \nonumber \\
  &&\s    -     \frac{N+2}{8 \pi^2}  {\tr}  [F(\widetilde{A}_\rmf)-B^{(2)}_\rmf]^2  \left[\diff A_2^{(1)}-{N\over 2}B^{(2)}_\rmc-{N+4\over 2}B^{(2)}_\rmf\right]     \nonumber\\
  && \s -  \frac{1}{8\pi^2}   \left(\frac{N}{2}\right)^2  \frac{(N+4)(N+3)}{2}     \left[{\diff} A+B^{(2)}_\rmc+B^{(2)}_\rmf\right]^2    \left[\diff A_2^{(1)}-{N\over 2}B^{(2)}_\rmc-{N+4\over 2}B^{(2)}_\rmf\right]   \nonumber \\
&&\s    -       \frac{1}{8\pi^2}  \frac{N}{2} \frac{(N+4)(N+3)}{2}    \left[{\diff} A+B^{(2)}_\rmc+B^{(2)}_\rmf\right]    \left[\diff A_2^{(1)}-{N\over 2}B^{(2)}_\rmc-{N+4\over 2}B^{(2)}_\rmf\right]^2   \nonumber \\   
  && \s  -  {1 \over 24\pi^2} \frac{(N+4)(N+3)}{2} \left[\diff A_2^{(1)}-{N\over 2}B^{(2)}_\rmc-{N+4\over 2}B^{(2)}_\rmf\right]^3   \;.     \label{Baryons}
      \eea
            Note that the second-from-the-last term,  corresponding to   $ [{\mathbbm Z}_2]^2 -U(1)_{\psi\eta}$ anomaly, is identical in  the UV  and in the IR, see Eq.~(\ref{UVanomaly}) and Eq.~(\ref{Baryons}).

\subsection{An almost flat  $({\mathbbm Z}_2)_F$ connection, generalized cocycle condition, and the 't Hooft fluxes }

Before proceeding to the actual determination of various mixed anomalies,  let us recapitulate some formal points involved in our analysis.  The first is 
 the meaning of the gauge field for $({\mathbbm Z}_2)_F$  introduced above. 
 The combination 
 \be       2   A_2^{(1)} -   B^{(1)}_\rmc -  B^{(1)}_\rmf     =    d   A_2^{(0)}\;,    \label{both}
\ee
 is the modification of the   $({\mathbbm Z}_2)_F$ gauge field,  
 $2   A_2^{(1)} = d   A_2^{(0)}$,   
such that it is invariant under the 1-form gauge transformations, (\ref{ZN1FG})-(\ref{werequire}).  
 By taking the derivatives of the both sides of  Eq.~(\ref{both})  it might appear that one gets 
\be     2   \,   \diff   A_2^{(1)}-{N}B^{(2)}_\rmc-  (N+4) B^{(2)}_\rmf  =0 \;:   \label{this}
\ee
this would erase all terms  containing  $  2\,\diff   A_2^{(1)}-{N}B^{(2)}_\rmc-  (N+4) B^{(2)}_\rmf $ from the $6D$ action, (\ref{psianomaly})-(\ref{Baryons}).
This, of course,  is not correct  as  $ A_2^{(0)}$ is a $2\pi$ periodic (angular) field. 
Indeed, the left hand side of   Eq.~(\ref{both})     is "an almost flat connection":   Eq.~(\ref{this}) is correct locally, but cannot be  set to zero
identically, as it can give nontrivial contribution when integrated over  $\Sigma_2$.

Actually,   by integrating the both sides of   Eq.~(\ref{both}) over a noncontractible cycle, one gets 
\bea   
&&   \oint dx^{\mu}    \big( 2   A_2^{(1)} -   B^{(1)}_\rmc -  B^{(1)}_\rmf  \big)_{\mu}     =   \oint    d   A_2^{(0)}  =   2\pi n\;,   \qquad  n\in {\mathbbm Z}\;,  \nonumber \\
&&   \oint   A_2^{(1)}  = \frac{ 2\pi m }{2} \;,   \qquad     m\in {\mathbbm Z}\;, \label{Z2gauge}
\eea
and
\be    \int_{\Sigma_2} N\, B^{(2)}_\rmc  +    \int_{\Sigma_2}  (N+4) \, B^{(2)}_\rmf  =      2\pi k\;,\qquad  k \in {\mathbbm Z}\;,  \label{consistentwith}
\ee
where $\Sigma_2$  is taken to be a nontrivial closed two-dimensional surface.
Eq.~(\ref{Z2gauge}) is a trademark of a $\Z_2$ gauge field. 
Eq.~(\ref{consistentwith})  is consistent with mutually independent  fluxes of  $B^{(2)}_\rmc$ and  $B^{(2)}_\rmf$,   (\ref{Byconstruction2})-(\ref{Byconstruction4}).   In passing, we note that this is in line with the remark made in Sec.~\ref{sec:mixedanomalies},  that the whole analysis of this
work could have been done possibly by keeping only one of  the  2-form gauge fields,  $B^{(2)}_\rmc$   or $B^{(2)}_\rmf$.

All this can be rephrased in terms of the generalized cocycle. In the case of standard QCD with massless left-handed and right-handed quarks,  the relevant symmetry  involves
 ${\mathbbm Z}_N\subset SU(N)_{\rmc}$ and   ${\mathbbm Z}_N \subset  U(1)_V$.  By compensating the failure of the cocycle condition at a triple overlap region of spacetime manifold   by a  color  ${\mathbbm Z}_N$ factor \footnote{In pure $SU(N)$ theory this would not be a problem, as the gauge fields do not feel the  ${\mathbbm Z}_N$  transformation:  it corresponds to the well-known statement that the pure $SU(N)$ theory  (or a theory with matter fields in adjoint representation)  is really an $\frac{SU(N)}{{\mathbbm Z}_N}$ gauge theory.  Alternatively, one can introduce nontrivial 
 't Hooft fluxes by introducing doubly periodic conditions with nontrivial  ${\mathbbm Z}_N$ twists.} by a simultaneous  ${\mathbbm Z}_N \subset  U(1)_V$ transformation,  one can formulate a consistent    $\frac{SU(N)}{{\mathbbm Z}_N}$ "QCD".\footnote{This has been worked out explicitly in \cite{Tanizaki}, Sec. 2.3.}

In our case, the failure of the straightforward cocycle condition by color ${\mathbbm Z}_N$  center factor can be compensated by 
a simultaneous   ${\mathbbm Z}_N \subset  U(1)_{\psi\eta} \times {\mathbbm Z}_2$  phase transformation of the fermions.  See Eq.~(\ref{ZNpsieta}) and 
Eq.~(\ref{ZNequiv}).   Similarly for the 
${\mathbbm Z}_{N+4}$ center.\footnote{In fact, this is the content of the $1$-form gauge invariance we impose. Eqs.~(\ref{ZN1FG})-(\ref{werequire})
can be regarded as the local form of the conditions, (\ref{eliminate1})-(\ref{eliminate2}) below.}    The 
consistency  for $\psi$ and  $\eta$  gauge transformations in a triple overlapping region thus reads
    \be    \left(e^{\im  \tfrac{2\pi}{N} n_{ij}} \right)^2 =  \mp   e^{- \im \tfrac{N+4}{2} \Delta \alpha_{ij}}\;, \label{eliminate1}
    \ee
    and
     \be   \left(e^{\im  \tfrac{2\pi}{N} n_{ij}} \right)^{-1}  e^{\tfrac{2\pi \im}{N+4} m_{ij}}      =  \mp    e^{\im \tfrac{N+2}{2} \Delta \alpha_{ij}}\;, \label{eliminate2}
    \ee
    respectively.  Finding   $e^{\im \Delta \alpha_{ij}}$ by multiplying   Eq.~(\ref{eliminate1}) and  Eq.~(\ref{eliminate2}) and inserting  it back, one 
gets a consistency condition
\be  e^{ \pi \im (n_{ij}+ m_{ij})}   = \mp 1 \;.
 \label{thiscondition}
\ee

 If   (\ref{eliminate1}) and  (\ref{eliminate2})  were to be interpreted in terms of  't Hooft's  twisted periodic conditions, the exponents in these formulas, 
 $\tfrac{2\pi}{N} n_{ij}$,     $\tfrac{2\pi}{N+4} m_{ij}$, $\pm \pi$,  $\Delta \alpha_{ij}$ would respectively be the $SU(N)$, $SU(N+4)$, $({\mathbbm Z}_2)_F$ and $U_{\psi\eta}(1)$
 fluxes through a closed two dimensional surface, $\Sigma_2$.  This requires some care, because of the discrete periodicity of the $({\mathbbm Z}_2)_F$ gauge field. In particular the presence of such  a flux means that, if  $\Sigma_2$ is taken as a torus,
there should be a point-like singularity on it  (2-dimensional surfaces, from the point of view of the four dimensional spacetime), carrying a  $(\mathbbm Z_2)_F$ flux. 
 Actually, it seems to us more natural, in the presence of a  $(\mathbbm Z_2)_F$ gauge field, to take as  $\Sigma_2$ not a torus with a singularity, but a smooth Riemann surface of genus $2$ (a double torus). 
 
%


   As already noted in Sec.~\ref{sec:mixedanomalies},  these indices  
$n_{ij}$  (or   $m_{ij}$)  correspond exactly to the second Stiefel-Whitney class of $\frac{SU(N)}{{\mathbbm Z}_N}$   (or
 $\frac{SU(N+4)}{{\mathbbm Z}_{N+4}}$) connections.  In other words,   the condition (\ref{thiscondition}) translates into the 
   $B^{(2)}_\rmc$ and  $B^{(2)}_\rmf$  flux relation,  Eq.~(\ref{consistentwith}).

 \subsection{Anomaly matching without the gauging of the 1-form center symmetries  \label{sec:oldanommatching} }

As another little preparation for our calculations,  let us first check that our gauge fields and their variations are properly normalized,  by considering   the anomalies in the ordinary case,   i.e.,  where  the 1-form  
${\mathbbm Z}_N$ and ${\mathbbm Z}_{N+4}$ symmetries are not gauged.  In other words, we set   
\be   B^{(2)}_\rmc = B^{(1)}_\rmc = B^{(2)}_\rmf= B^{(1)}_\rmf= 0\;. \ee 
The first three terms  (the triangles involving  $U(1)_{\psi\eta}$ and $SU(N+4)$)  of Eq.~(\ref{Baryons}) match  exactly those in the UV anomaly,   Eq.~(\ref{UVanomaly}), whether or not $ \big(B^{(2)}_\rmc\,, B^{(2)}_\rmf \big)$ fields are present.  
The second-from-the-last terms in Eq.~(\ref{UVanomaly}) and in Eq.~(\ref{Baryons})  describe the  nontrivial   $[({\mathbbm Z}_2)_F]^2 -  U(1)_{\psi\eta}$  anomaly,  which are identical  in UV and IR,    again,  whether or not the 1-form gauging of  ${\mathbbm Z}_N$ and ${\mathbbm Z}_{N+4}$  is done.

  To compute  the  $({\mathbbm Z}_2)_F$  anomaly  in the  UV,   one collects the terms 
  \be    \int_{\Sigma_6}  (\ldots)  \,  \diff A_2^{(1)}\;, 
  \ee  
  and integrate to get the boundary $5D$ effective WZW  action
  \be    \int_{\Sigma_5}     (\ldots)   \,  A_2^{(1)}\;.   
  \ee
  The $({\mathbbm Z}_2)_F$   transformations of the fermions are  formally  expressed as  the transformation of the $({\mathbbm Z}_2)_F$ "gauge field"   $A_2^{(1)}$,
  \be   A_2^{(1)}  \to  A_2^{(1)}   +   \frac{1}{2}\,    d  (\delta  A_2^{(0)})\;,   \qquad   \delta  A_2^{(0)} = \pm 2  \pi\;,
  \ee  
  yielding the  anomaly-inflow   in $4D$ 
 \bea 
      \delta S^4_{\rm UV}   &=&    -     \frac{2}{8 \pi^2} \int_{\Sigma_4}  {\tr}  [F({a})]^2  \, \frac{\delta A_2^{(0)}}{2}    -     \frac{N}{8 \pi^2}  \int_{\Sigma_4} {\tr}  [F({A}_\rmf)]^2   \,\frac{\delta A_2^{(0)}}{2}   \nonumber 
      \\
&& -  \frac{1}{8\pi^2} \frac{N(N+4)(N^2+3N+4)}{8}  \int_{\Sigma_4}     \left[{\diff} A \right]^2   \, \frac{\delta A_2^{(0)}}{2}  \;.
     \label{UVanomalyno1form}
\eea  
The first line is the standard  chiral anomaly expression associated with the field transformation
\be  \psi \to - \psi\;, \qquad  \eta \to - \eta\;,   \qquad      \delta A_2^{(0)} =  \pm 2  \pi\;
\ee
  due to $SU(N)$ and $SU(N+4)$  gauge fields.    They are actually both trivial (no anomalies) due to the integer instanton numbers:
\be      \frac{1}{8\pi^2}    \int_{\Sigma_4}     {\tr}  [ F(a) ]^2 = \mathbbm Z\;,  \qquad    \frac{1}{8\pi^2}    \int_{\Sigma_4}     {\tr} [F({A}_\rmf)]^2 = \mathbbm Z\;.    \label{indeed}
\ee
Note that, crucially,  their coefficients ($2$ and $N$) are both even integers. This confirms that the field  $A_2^{(1)}$ and its variation
$\delta A_2^{(0)} $ are correctly  normalized. 

Similarly in the IR one has  
\bea 
 \Delta S^4_{\rm IR}   & = &    -     \frac{N+2}{8 \pi^2} \int  {\tr}  [F({A}_\rmf)]^2      \,\frac{\delta A_2^{(0)}}{2}  \nonumber \\
&& -  \frac{1}{8\pi^2}   \left(\frac{N}{2}\right)^2  \frac{(N+4)(N+3)}{2}   \int  \left[{\diff} A \right]^2    \,\frac{\delta A_2^{(0)}}{2}   \;.
     \label{IRanomaly4Dno1form}
\eea  
Again, the first term is trivial, as $N+2$ is an even integer.

The second terms in Eq.~(\ref{UVanomalyno1form}) and in Eq.~(\ref{IRanomaly4Dno1form})  describe the  nontrivial   $({\mathbbm Z}_2)_F -  [U(1)_{\psi\eta}]^2$  anomaly,
present  both in the UV and in the IR.\footnote{This is so for even $N$ of the form,  $N= 4m +2$,    $m \in {\mathbbm Z}$.}    However, their difference  is given by 
\be   -   \frac{ N (N+4)}{2}        \int  \left(  \frac{1}{8\pi^2}   \left[{\diff} A\right]^2   \right)     \cdot     \,\frac{\delta A_2^{(0)}}{2} \;.
\ee
Since the coefficient $\frac{N (N+4)}{2} $  is any even integer   the discrete $({\mathbbm Z}_2)_F-  [U(1)_{\psi\eta}]^2$
anomaly is matched modulo ${\mathbbm Z}_2$ in the IR and UV.

All in all,  we  reproduce the earlier results reported in Sec.~\ref{sec:earlier}, that a chirally symmetric vacuum, with no condensates, with no NG bosons but 
with massless baryons   Eq.~(\ref{baryons10}),    satisfy  all the conventional   't Hooft anomaly matching constraints.

      \section{UV-IR matching of various mixed anomalies in even $N$ theories \label{sec:computing}}

   Now we come to the main issues of our analysis:  studying the various mixed anomalies involving  the fermion parity $({\mathbbm Z}_2)_F$, in the presence of
      the 2-form gauge fields  $B^{(2)}_\rmc$ and $B^{(2)}_\rmf$, in an even $N$ theory.  
         Starting from the $6D$ action,  Eq.~(\ref{psianomaly}) - Eq.~(\ref{Baryons}),     one collects the terms of the form,  
 \be    S^{6D}=     \int_6   \, [...] \, \left[\diff A_2^{(1)}-{N\over 2}B^{(2)}_\rmc-{N+4\over 2}B^{(2)}_\rmf\right]   \;.
 \ee
Integrating,  one gets the 5D boundary WZW action 
 \be   S^{5D} =       \int_5  \,   [...]  \,\left[A_2^{(1)}-\frac{1}{2}B^{(1)}_\rmc- \frac{1}{2} B^{(1)}_\rmf\right] \;.
 \ee
 This allows us to calculate various  anomalies in $4D$  involving   $({\mathbbm Z}_2)_F$,    by  anomaly inflow,  considering the variations 
 \be      \delta    [ A_2^{(1)}-\frac{1}{2}B^{(1)}_\rmc- \frac{1}{2} B^{(1)}_\rmf ]  =     \frac{1}{2}  \,  \diff  \, \delta A_2^{(0)}     \;,
 \ee
 \be   \delta  S^{4D}   =   \frac{1}{2}    \int_4   \,  [...]   \,   \delta A_2^{(0)}\;,    \qquad  \delta A_2^{(0)}  = \pm 2\pi\,.
 \ee

      \subsection{Mixed  $ ({\mathbb Z}_{2})_{F}- [\mathbb{Z}_N]^2 $ anomaly  \label{mostinteresting}}
     Collecting all terms of the form
      \be    N^2   \int  (B^{(2)}_\rmc)^2   A_2^{(1)}    
      \ee
      in  the $5D$   WZW action,   one finds at UV,
      \be           S^{(5)}_{\rm UV}   =     1\cdot    \frac{1}{8\pi^2}  \int_{\Sigma_5}   N^2    (B^{(2)}_\rmc)^2   \cdot    A_2^{(1)}    \;. \label{Z2Anomaly} 
      \ee  
      The coefficient in front of the above expression (\ref{Z2Anomaly})   is the result of the sum from various  $\psi$ and $\eta$   contributions
      in (\ref{psianomaly}) and (\ref{etaanomaly}):
       \bea  && - \frac{N+2}{N} + \frac{(N+1)(N+4)^2}{8N} - \frac{(N+4)(N+1)}{4} +  \frac{N(N+1)}{8}  \nonumber \\
   &&   + \frac{N+4}{N}  -  \frac{(N+4)(N+2)^2}{4N} +  \frac{(N+2)(N+4)}{2} - \frac{N(N+4)}{4} =1\;.   \label{correctly} 
      \eea
     The result (\ref{Z2Anomaly}) leads to the $4D$ mixed $ ({\mathbb Z}_{2})_{F}- [\mathbb{Z}_N]^2 $ anomaly  in the UV, 
     \be     \Delta  S^{(4)}_{\rm UV}   =     \pm \im \pi   {\mathbbm Z}\;,\qquad    \delta A_2^{(1)}  =  d \,      \frac{1}{2}     \delta  A_2^{(0)}\;, \qquad      \delta   A_2^{(0)}  =\pm  2\pi\,.
     \ee
    In other words, the partition function changes sign under the ${\mathbbm Z}_2$ transformation, $ \psi \to -\psi$,  $ \eta \to -\eta$, 
      in appropriate  background $B^{(2)}_\rmc$ fields \footnote{Equivalently, in the presence of appropriate fractional 't Hooft fluxes. } 
     \be   \frac{1}{8\pi^2}  \int_{\Sigma_4}   N^2    (B^{(2)}_\rmc)^2 ={\mathbbm Z}\;.
      \ee

On the other hand,  in the infrared,  assuming the chirally symmetric  scenario, Sec.~\ref{possible0},  the "massless baryons"  (\ref{Baryons})   lead to no anomalies of this type: 
       \be    0 \cdot   N^2    (B^{(2)}_\rmc)^2   A_2^{(1)}   = 0\;,    
      \ee
      due to the cancellation
      \be    -  \frac{(N+4)(N+3)}{8}  -  \frac{(N+4)(N+3)}{8} +  \frac{(N+3)(N+4)}{4} =0\;,
      \ee
      among the $4$th, $6$th and $7$th terms of  (\ref{Baryons}).   
Actually the absence of the mixed $A_2^{(1)} - B^{(2)}_\rmc$ terms can be seen directly  from the first line of  (\ref{Baryons}). 
(See footnote 13.)

 The conclusion is that  the   mixed   $ ({\mathbb Z}_{2})_{F}- [\mathbb{Z}_N]^2 $
 anomaly is present in the UV but absent in the IR. 
 They do not match.

         \subsection{Mixed  $ ({\mathbb Z}_{2})_{F}- [\mathbb{Z}_{N+4}]^2$  anomaly}
      
      We now  study the terms
       \be    (N+4)^2  \int   (B^{(2)}_\rmf)^2   A_2^{(1)}    
      \ee
       in the $5D$ action.  The $\psi$ and $\eta$ both give vanishing contribution to the coefficient: 
         \bea  && - \frac{N(N+1)}{8} +   \frac{N(N+1)}{4} - \frac{N(N+1)}{8}  =0\;,  \nonumber \\
   &&   \frac{N}{N+4}  -    \frac{N (N+2)^2}{4(N+4)} +  \frac{N(N+2)}{2} - \frac{N(N+4)}{4} =   0\;.
      \eea
         On the other hand,  the massless baryons in the IR   gives:
      \be    -   \frac{(N+3)(N+4)}{8} + \frac{N+2}{N+4} -   \frac{N^2 (N+3)}{8 (N+4)} +   \frac{N(N+3)}{4}  =  -1\;. 
      \ee
Therefore,  the result  here  is  opposite:   the   mixed  $ ({\mathbb Z}_{2})_{F}- [\mathbb{Z}_{N+4}]^2$   anomaly is absent  in the UV but present  in the IR! 
 However the conclusion is the same:   they do not satisfy the 't Hooft anomaly  matching requirement.

            \subsection{Mixed  $({\mathbbm Z}_2)_F -  {\mathbbm Z}_N- {\mathbbm Z}_{N+4}$    anomaly}
         
      Let us now consider the mixed anomalies of the type,    $({\mathbbm Z}_2)_F -  {\mathbbm Z}_N- {\mathbbm Z}_{N+4}$.  
      We collect the terms of the form  
       \be    N  (N+4)  \int B^{(2)}_\rmc  B^{(2)}_\rmf   A_2^{(1)}    
      \ee
      in the $5D$ action.  The result in the UV is that  $\psi$  gives the coefficient 
      \be        \frac{(N+4)(N+1)}{4} -  \frac{N+1}{4} \cdot (2N+4)  +   \frac{N(N+1)}{4}=0\;,
      \ee
      whereas  $\eta$ yields  
      \be      -  \left( \frac{N+2}{2}\right)^2 \cdot 2 +    \frac{N+2}{2}   (2N+4)  -   N(N+4)  \frac{1}{2}   =2\;.
      \ee
      Thus there are no  mixed  $({\mathbbm Z}_2)_F -  {\mathbbm Z}_N- {\mathbbm Z}_{N+4}$    anomaly in the UV.

      In the IR, the baryons produces the terms of this type with the coefficient: 
      \be     -  \frac{(N+3)(N+4)}{2}  \frac{1}{2} -    \frac{N(N+3)}{4}  +   \frac{ N+3}{4}   (2N+4)   =0\;. 
      \ee  
      (Again this result could have been read off from the first line of (\ref{Baryons}).)
      Therefore there are no anomalies of this type in the IR either.  Therefore no question of 't Hooft consistency condition arises 
      from the consideration of the mixed     $({\mathbbm Z}_2)_F -  {\mathbbm Z}_N -  {\mathbbm Z}_{N+4}$    anomalies. 
    
       \subsection{Mixed  $({\mathbbm Z}_2)_F -  {\mathbbm Z}_N- U(1)_{\psi\eta}$    anomaly}

       In the UV, one collects the terms of the form, 
        \be       N  \int   B^{(2)}_\rmc    dA   \,   A_2^{(1)}    
      \ee
      in the $5D$ action. One finds the coefficient,
      \bea   &&  \left( \frac{N+4}{2}\right)^2  (N+1) -   \frac{N+4}{2}  \frac{N(N+1)}{2}  -     \left( \frac{N+2}{2}\right)^2  2 (N+4) + \frac{N+2}{2} N(N+4) \nonumber \\
      && =
      - N-4\;,
      \eea
       which is an even integer. This means that no mixed  $({\mathbbm Z}_2)_F -  {\mathbbm Z}_N- U(1)_{\psi\eta}$ anomaly is present  in the UV. 
       We know already that there are no terms mixing   $ A_2^{(1)}  $ and  $B^{(2)}_\rmc $  in the infrared:   there are no mixed anomalies of this type in the infrared either.

         \subsection{Mixed  $({\mathbbm Z}_2)_F -  {\mathbbm Z}_{N+4} - U(1)_{\psi\eta}$    anomaly}

One must collect the terms of the form, 
        \be       (N+4)  \int   B^{(2)}_\rmf    dA   \,   A_2^{(1)}    \;.
      \ee
      One finds the coefficients, in the UV, 
      \bea  && \psi \;: \quad    \frac{N+4}{4}  N (N+1)  -  \frac{1}{2}  \frac{N(N+1)}{2} (N+4) =0\;,   \nonumber \\   
        &&   \eta \;:   \quad  - 2 N  \left(\frac{N+2}{2}\right)^2  +  \frac{N+2}{2} N (N+4)  =  N(N+2)\;,
         \eea 
     the sum of which is an even integer: there are no anomaly of this type in the UV.     In the IR, the massless baryons give
      \be   - \left(\frac{N}{2}\right)^2  (N+3)  +   \frac{N}{2}  \frac{(N+3)(N+4)}{2} = N(N+3)\;,
      \ee
      which is again an even integer.  There are no anomaly of this type  in the IR either.

   \subsection{Physics implications \label{sec:implication}} 

Of all types of mixed anomalies involving  the fermion parity   $({\mathbbm Z}_2)_F$ considered above,   we thus find that  
 $ ({\mathbb Z}_{2})_{F}- [\mathbb{Z}_N]^2 $  and   $ ({\mathbb Z}_{2})_{F}- [\mathbb{Z}_{N+4}]^2$  anomalies provide us with the most interesting information.
 Namely the anomaly of the first type  is present in the UV but absent in the IR;  the situation is opposite for the second type of anomaly: it is 
absent in the UV but present in the IR.    All other types of mixed anomalies as well as conventional anomalies are found to match in the UV and IR, 
assuming the chirally symmetric vacuum of Sec.~\ref{possible0}.
 
   We are thus led to conclude that the chirally symmetric vacuum of Sec.~\ref{possible0}  cannot be the correct vacuum of the $\psi\eta$   theory with even $N$. 
  
  No problem arises if the system is in the dynamical Higgs phase,   discussed in  Sec.~\ref{possible1}. One might however  wonder how the failure of the matching of these mixed anomalies in 
the UV and IR might be accounted for by the bifermion condensate,   $\brc \psi \eta \ckt$,  in view of the fact that  the fermion parity   ($2\pi$ space rotation) does not act on it.  
The answer is that the failure of the 't Hooft matching condition in this case means that  
 the 1-form gauging of the $[U(1)_{\psi\eta} \times ({\mathbbm Z}_2)_F -SU(N)]$-locked   $\mathbb{Z}_N$, and  the $[U(1)_{\psi\eta}  \times ({\mathbbm Z}_2)_F -SU(N+4)]$-locked  $\mathbb{Z}_{N+4}$, 
 center symmetries
is not allowed.  The condensates $\brc \psi \eta \ckt$  indeed breaks spontaneously both of the global 0-form $U(1)_{\psi\eta}$ and  the global 1-form    $\mathbb{Z}_N$   color center   (or  the flavor $\mathbb{Z}_{N+4}$  center) symmetry, the infrared system being in a dynamically induced  Higgs phase. 

Still, a little more careful argument is necessary, before jumping to the conclusion that everything is consistent in the Higgs phase.   The reason is that  the  $\brc \psi \eta \ckt$ condensates leaves a nontrivial subgroup (\ref{symmHig}) unbroken, and  that some massless fermions are present in the IR  so that the conventional perturbative anomaly matching works. This means that the {\it generalized} anomaly matching requirement (in the presence of some combination of the 2-form gauge fields $\big( B^{(2)}_\rmc, B^{(2)}_\rmf \big)$ appropriate for the 
unbroken symmetry group (\ref{symmHig}))
might fail to be satisfied in the dynamical Higgs phase, too.   

Actually, an attentive inspection of Table~\ref{SimplestBis} dispels  the last worry.  When the Dirac pair  of massive fermions
($\psi$ and  the symmetric part in  ${\tilde \eta}^A_i$) are excluded, the rest of the massless fermions in the UV are identical to the set of the massless "baryons" in the IR,  in all their quantum numbers,  charges, and multiplicities. This means whatever subset of  $\big( B^{(2)}_\rmc, B^{(2)}_\rmf \big)$ are retained,
the UV-IR matching is automatically satisfied.

 \section{Calculating the mixed anomalies without  Stora-Zumino  \label{sec:withoutSZ}} 
 
 In the above we made use of the Stora-Zumino descent method to calculate the various anomaly expressions. 
It has a great advantage of being systematic, yielding the 
Abelian, nonAbelian and other, mixed types of anomalies all at once with the correct coefficients,  and showing certain aspects of symmetries.
 Nevertheless, it is basically a technical aspect of our analysis:  it  is not indispensable. Indeed, one can stay in four-dimensional spacetime, and 
 calculate the anomalies of the chiral transformation,  $\psi\to -\psi$, $\eta\to -\eta$, in the underlying (UV)  theory, in the standard fashion, e.g. by Fujikawa's method \cite{Fujikawa}.  By taking into account the 1-form gauge invariance requirement, 
 Eq.~(\ref{ZN1FG}),  Eq.~(\ref{ZN41FG}),  Eq.~(\ref{werequire}), however,   one finds     ($\im  \pi$  times)
 \be      -     \frac{N+4-(N+2)}{8 \pi^2} \int_{\Sigma_4}  \, \,  {\tr}  \big(F(\tilde{a})-B^{(2)}_\rmc\big)^2  \;,   \label{modific1}
 \ee
 \be      -       \frac{N}{8 \pi^2}   \int_{\Sigma_4} \,   \,   {\tr}  [F(\widetilde{A}_\rmf)-B^{(2)}_\rmf]^2   \, \;,
 \ee
 \be  -  \frac{1}{8\pi^2} \frac{N(N+4)(N^2+3N+4)}{8}     \int_{\Sigma_4} \,       \left[{\diff} A+B^{(2)}_\rmc+B^{(2)}_\rmf\right]^2\;,  \ee
 \be   -       \frac{1}{8\pi^2}  \frac{N(N+3)(N+4)}{4}    \int_{\Sigma_4} \,      \left[{\diff} A+B^{(2)}_\rmc+B^{(2)}_\rmf\right]    \left[\diff A_2^{(1)}-{N\over 2}B^{(2)}_\rmc-{N+4\over 2}B^{(2)}_\rmf\right]      \;,  \label{modific4}
 \ee
 \be   -  {1 \over 24\pi^2} \frac{N(N+7)}{2}   \int_{\Sigma_4}  \,     \left[\diff A_2^{(1)}-{N\over 2}B^{(2)}_\rmc-{N+4\over 2}B^{(2)}_\rmf \right]^2\;,     \label{modific5}
 \ee
 due to the external fields,   
 \be     [SU(N)]^2\;,   \quad  [SU(N+4)]^2\;, \quad  [U(1)_{\psi\eta}]^2\;, \quad  U(1)_{\psi\eta}   {\mathbbm Z}_2\;,  \quad  
 [{\mathbbm Z}_2]^2\;,
 \ee
 dressed by the 2-form gauge fields   $\big( B^{(2)}_\rmc, B^{(2)}_\rmf \big)$,     respectively.\footnote{Eqs.~(\ref{modific1})-(\ref{modific5}) are obtained by
 taking $({\mathbbm Z}_2)_F$ to be  $\psi \to e^{\im\pi} \psi;$    $\eta \to e^{- \im\pi} \eta$, in accordance with the convention used in  Eqs.~(\ref{steps})-(\ref{etastrength}).  If the  phase of $\eta$ were  to be  chosen as  $+\im \pi$,  the coefficients in Eqs.~(\ref{modific4}) and (\ref{modific5}) will get modified, but the final  result for $({\mathbbm Z}_2)_F$ anomaly  remains unchanged.  }   A similar consideration can be made for the calculation of anomaly in the IR. 
 Collecting various terms of the same types, one ends up with the results presented  in Sec.~\ref{sec:computing}.

 It might be of interest  to recall a subtle aspect  in the descent procedure, noted after Eq.~(\ref{etastrength}).   In a $4D$ calculation described here,  it is   manifest  that we are modifying our theory,  in going from the original $SU(N)\times SU(N+4)$ gauge theory to  $\frac{SU(N)}{{\mathbbm Z}_N}   \times \frac{SU(N+4)}{{\mathbbm Z}_{N+4}}$  theory.

 \section{Summary and discussion \label{sec:summary}}  
 
To summarize, in this note we have examined the symmetries of  a simple chiral gauge theory, $SU(N)$   $\psi \eta$    model,
by use of the recently  found extension of  the  't Hooft anomaly matching constraints,  to  include the mixed anomalies involving  some higher-form symmetries
(in our case,  some 1-form center symmetries).    A particular interest in this model lies in the fact that  the conventional 't Hooft anomaly matching constraints  allow a chirally symmetric confining vacuum,  with no condensates breaking the  $U(1)_{\psi\eta}\times SU(N+4)$ flavor symmetries, and with a set of massless baryonlike composite fermions saturating all the anomaly triangles.  Another possible type of vacuum, compatible with the anomaly matching conditions, is in a dynamical Higgs phase, with a bifermion condensates breaking color completely, but leaving some residual flavor symmetry.   The standard anomaly matching constraints do not tell apart the two possible dynamical possibilities, which represent two  distinct phases of the theory. 

The result of our investigation is that, a deeper level of consistency requirement, taking into account  also certain possible mixed  (0-form$-$1-form) anomalies,  allows  us, for  even $N$ theory at least,  to exclude the first, chirally symmetric type of vacua.   One is led inevitably to the conclusion that the 
system is likely to be in a dynamical Higgs phase.

More concretely,  among all possible  mixed anomalies involving  the  $({\mathbbm Z}_2)_F$  symmetry of the system,  which  corresponds actually to $2\pi$ space rotation,   the anomalies of the types $ ({\mathbb Z}_{2})_{F}- [\mathbb{Z}_N]^2 $  and   $ ({\mathbb Z}_{2})_{F}- [\mathbb{Z}_{N+4}]^2$
are present, and do not match in the UV and in the IR,  if the chirally symmetric vacuum is assumed.  

Our extension of the idea of gauging 1-form center symmetries  such as  ${\mathbbm Z}_N \subset SU(N)$ and of finding possible associated mixed anomalies,  as compared to the existent literature \cite{GKSW}-\cite{BKL},  involves a few new concepts.   Thus it may be useful to summarize them.   The first concerns the fact that the presence of fermions in the fundamental representation of the color $SU(N)$ (or of  the flavor $ SU(N+4)$) group, requires us to work with color-flavor locked center symmetries, see Eq.~(\ref{ZNequiv}), Eq.~(\ref{ZN+4equiv}). This involves the centers of the $SU(N)$  or  $SU(N+4)$  locked with some subgroups of the anomaly-free $U(1)_{\psi\eta}$.  A similar idea  has been studied and tested in several papers already, see \cite{ShiYon,TanKikMisSak,Tanizaki}.

The second nontrivial conceptual extension  here involves the discrete  $({\mathbbm Z}_2)_F$ symmetry for even $N$ theory.  In this case the center symmetry of interest is the diagonal combination of  ${\mathbbm Z}_N\subset SU(N)$ and   ${\mathbbm Z}_N \subset    U(1)_{\psi\eta} \times ({\mathbbm Z}_2)_F$.
 Similarly for  ${\mathbbm Z}_{N+4}$.   This means that both $U(1)_{\psi\eta}$  and $({\mathbbm Z}_2)_F$  gauge fields
transform nontrivially under the (gauged)  center symmetries, see  Eqs.~(\ref{ZN1FG})-(\ref{werequire}).  

From the  formal point of view, therefore,  the position of  $U(1)_{\psi\eta}$  and $({\mathbbm Z}_2)_F$ symmetries  (hence of the associated
background gauge fields) is therefore similar. Even though these are both 0-form symmetries
they carry charges under the gauged center  ${\mathbbm Z}_N$ or  ${\mathbbm Z}_{N+4}$ symmetry.  The anomalies involving  $U(1)_{\psi\eta}$  and $({\mathbbm Z}_2)_F$ are both modified nontrivially by the presence of the 2-form gauge fields,   $\big( B^{(2)}_\rmc, B^{(2)}_\rmf \big)$. 

There is an important difference,  however.  In the case of the continuous  $SU(N+4)\times U(1)_{\psi\eta}$  symmetries,  the anomaly triangles were
all matched in the UV and IR  {\it before} the introduction of $\big(B^{(2)}_\rmc, B^{(2)}_\rmf\big)$.  For instance,  the $[U(1)_{\psi\eta}]^3$ anomaly takes the simple form in the $6D$ action, $C\, ({\diff} A)^3$.  The anomaly coefficients satisfy, in the chirally symmetric vacuum  of Sec.~\ref{possible0}, the matching condition,
\be    C_{\rm UV}= C_{\rm IR}\;. \label{sufficient}  \ee
Now the introduction of the 2-form gauge fields   $\big( B^{(2)}_\rmc, B^{(2)}_\rmf \big)$  modifies all the fields,   e.g.,   
 \be   {\diff} A     \to   {\diff} A+B^{(2)}_\rmc+B^{(2)}_\rmf\;, \qquad  \diff A_2^{(1)} \to   \diff A_2^{(1)}-{N\over 2}B^{(2)}_\rmc-{N+4\over 2}B^{(2)}_\rmf\;,
 \ee
 etc.,   but clearly the matching condition (\ref{sufficient}) for the conventional  $U_{\psi\eta}(1)^3$ anomaly is sufficient to guarantee automatically  the
 matching of the anomaly 
 \be        C\, ({\diff} A+B^{(2)}_\rmc+B^{(2)}_\rmf)^3\;,
 \ee
 in the modified theory.
 The same applies to all triangle anomalies involving  the continuous  $SU(N+4)\times U(1)_{\psi\eta}$ symmetries.
 
 It is a different story for  the anomalies involving the discrete symmetry  $({\mathbbm Z}_2)_F$. Before the introduction of $\big(B^{(2)}_\rmc, B^{(2)}_\rmf\big)$,  
$({\mathbbm Z}_2)_F$ was a nonanomalous symmetry of the system. But this was so due to the integer instanton numbers, not because of an algebraic cancellation
between the contributions from different fermions, as for $U_{\psi\eta}(1)$.   Also, the $({\mathbbm Z}_2)_F$ anomaly "matching"  was not  due to the equality of the coefficients 
as in (\ref{sufficient}), but only  due to an equality {\it modulo} ${\mathbbm Z}_2$ of the coefficients, {\it and} under the assumption of integer instanton numbers   
\be \frac{1}{8\pi^2}    \int_{\Sigma_4}   \, F^2  \in {\mathbbm Z}\;. \label{paragraph}
\ee
This means that the introduction of the 2-form gauge fields (which can introduce nontrivial 't Hooft fluxes, hence fractional instanton numbers)  may make it anomalous,  and as a consequence may invalidate  the discrete anomaly matching.  Our calculation shows that it indeed does.

The result found here is 
somewhat reminiscent of the fate of the  time reversal  (or CP)  symmetry in the infrared,  
in pure  $SU(N)$ YM  theory with  $\theta=\pi$ \cite{GKKS}. 
 Note that before introducing the  ${\mathbbm Z}_N$  1-form gauging,  time reversal invariance at  $\theta=\pi$
holds because of the integer instanton numbers,  just as the  fermion parity symmetry $({\mathbbm Z}_2)_F$ of our system. 
From this prospect, what is found  here, $ ({\mathbb Z}_{2})_{F}- [\mathbb{Z}_N]^2 $   and $ ({\mathbb Z}_{2})_{F}- [\mathbb{Z}_{N+4}]^2 $   mixed anomalies, 
are very much  analogous  to the time reversal - 1-form  ${\mathbbm Z}_N$ mixed anomaly  discovered in the pure YM at $\theta=\pi$.
  Here the time reversal (CP symmetry) is replaced by $2\pi$ space rotation. 
 
Note however that  the way  the failure of the 't Hooft anomaly matching is reflected in the infrared physics is different here from the CP invariance
for the pure YM  at  $\theta=\pi$. In the latter case, a double vacuum degeneracy and the spontaneous breaking of CP in the infrared
"take care" of the eventual inconsistency which would arise  if we were to  gauge the 1-form ${\mathbbm Z}_N$ center  symmetry.  

Here, the failure  of the  mixed-anomaly matching is "accounted for" in the infrared, dynamically Higgsed phase,  not by the spontaneous breaking of the fermion parity symmetry, but by the  breaking of the 1-form   ${\mathbbm Z}_N$  and ${\mathbbm Z}_{N+4}$ symmetries.  
Note that in our system, the   1-form   symmetries are locked with 
 $U(1)_{\psi\eta}$ symmetry, which  is  spontaneously broken by the bifermion condensate  $\brc \psi \eta \ckt$.   
It is true, as noted at the end of Sec.~\ref{sec:computing}, that some subgroup of the original symmetry group with nontrivial global structure 
(\ref{symmHig}) survives the bifermion condensates. But as noted also there, the eventual anomalies with respect to the surviving symmetries
match completely in the UV and in the IR, in the case of the dynamical Higgs phase.  The hypothesis of gauge noninvariant bifermion condensate  (\ref{cflocking})
 is therefore consistent with our symmetry arguments.  On the contrary, the chirally symmetric vacuum contemplated earlier in the literature is, at least for even $N$,  inconsistent: it cannot be realized dynamically.

 \section*{Acknowledgments}
 
 The work  is supported by the INFN special research grant, "GAST (Gauge and String Theories)".
 We thank Yuya Tanizaki for the collaboration at the early stage of the work and for useful discussions.  K.K. thanks Erich Poppitz for prodding him to investigate the implications of the mixed anomalies in chiral gauge theories such as the ones studied here.

\end{document}